\documentclass[a4paper,12pt]{article}
\usepackage[dvips]{graphics}
\usepackage{epsfig}
\usepackage{psfrag}
\usepackage[psamsfonts]{amssymb}
\usepackage{amsmath}
\usepackage{indentfirst}
\usepackage{amssymb}
\usepackage{wrapfig}

\usepackage{graphicx}
\usepackage{amsmath}
\usepackage{amssymb}
\usepackage{theorem}
\usepackage{url}

\newcommand \be{\begin{equation}}
\newcommand \ba{\begin{eqnarray}}
\newcommand \ee{\end{equation}}
\newcommand \ea{\end{eqnarray}}

\newcommand{\cL}{{\cal L}}
\newcommand{\lgl}{\langle}
\newcommand{\rgl}{\rangle}

\begin{document}

\author{Didier Sornette\\
ETH Zurich\\
Scheuchzerstrasse 7, CH-8092 Zurich, Switzerland\\
dsornette@ethz.ch}

\rule{0mm}{20mm}

\centerline{\Large  Physics and Financial Economics (1776-2014)}
\smallskip
\centerline{\Large Puzzles, Ising  and Agent-Based models}

\rule{0mm}{5mm}

\centerline{\large D. Sornette$^{1,2}$}


\centerline{$^{1}$\footnotesize ETH Zurich -- Department of Management, Technology and Economics, 
Scheuchzerstrasse 7, CH-8092 Zurich, Switzerland}

\centerline{$^{2}$\footnotesize Swiss Finance Institute, 40, Boulevard du Pont-d' Arve, Case Postale~3, 1211 Geneva 4, Switzerland}

\centerline{\footnotesize E-mail address: dsornette@ethz.ch}

\begin{abstract}
This short review presents a selected
history of the mutual fertilization between physics and economics,
from Isaac Newton and Adam Smith to the present.
The fundamentally different perspectives
embraced in theories developed in financial economics compared with physics
are dissected with the examples of the volatility smile and of the excess volatility puzzle.
The role of the Ising model of phase transitions to model social and financial systems
is reviewed, with  the concepts of random utilities and the logit model
as  the analog of the Boltzmann factor in statistic physics. Recent extensions
in term of quantum decision theory are also covered.
A wealth of models are discussed briefly that build on the Ising model and generalize
it to account for the many stylized facts of financial markets.
A summary of the relevance of the Ising model and its extensions
is provided to account for financial bubbles and crashes. 
The review would be incomplete if it would not cover the dynamical
field of agent based models (ABMs), also known as computational economic models,
of which the Ising-type models are just
special ABM implementations. We formulate the ``Emerging Market Intelligence 
hypothesis'' to reconcile the pervasive presence of ``noise traders''
with the near efficiency of financial markets. Finally, we note that
evolutionary biology, more than physics, is now playing a growing role 
to inspire models of financial markets.
\end{abstract}

\vskip 0.5cm 

\noindent
Keywords:  Finance, physics, econophysics, Ising model, phase transitions, 
excess volatility puzzle, logit model, Boltzmann factor, bubbles, crashes, 
adaptive markets, ecologies

\vskip 0.5cm 
\noindent
JEL:   A12, B41, C00; C44; C60; C73; D70; G01

\clearpage

\tableofcontents

\clearpage

\section{Introduction}

The world economy is an extremely complex system with hidden causalities rooted in intensive social and technological developments. Critical events in such systems caused by endogenous instabilities can lead to huge crises wiping out of the wealth of whole nations. On the positive side, positive feedbacks of education and venture capital investing on entrepreneurship can weave a virtuous circle of great potential developments for the future generations. Risks, both on the downside as well as on the upside, are indeed permeating and controlling the outcome of all human activities and require high priority. 

Traditional economic theory is based on the assumptions of rationality of economic agents and of their homogeneous beliefs, or equivalently that their aggregate behaviors can be represented by a representative agent embodying their effective collective preferences. However, many empirical studies provide strong evidences on market agentsÕ heterogeneity and on the complexity of market interactions. Interactions between individual market agents for instance cause the order book dynamics, which aggregate into rich statistical regularities at the macroscopic level. In finance, there is growing evidence that equilibrium models and the efficient market hypothesis (EMH), see section \ref{gsnewh} for an extended presentation and generalisation, cannot provide a fully reliable framework for explaining the stylized facts of price formation (Fama, 1970). Doubts are further fuelled by studies in behavioral economics demonstrating limits to the hypothesis of full rationality for real human beings (as opposed to the homo economicus posited by standard economic theory). We believe that a complex systems approach to research is crucial to capture the inter-dependent and out-of-equilibrium nature of financial markets, whose total size amounts to at least 300\% of the world GDP and of the cumulative wealth of nations.

From the risk management point of view, it is now well established that the Value-at-Risk measure, on which prudential Basel I and II recommendations are based, constitutes a weak predictor of the losses during crises. Realized and implied volatilities as well as inter-dependencies between assets observed before the critical events are usually low, thus providing a completely misleading picture of the coming risks. New risk measures that are sensitive to global deteriorating economic and market conditions are yet to be fully developed for better risk management.

In todayÕs high-tech era, policy makers often use sophisticated computer models to explore the best strategies to solve current political and economic issues. However, these models are in general restricted to two classes: (i) empirical statistical methods that are fitted to past data and can successfully be extrapolated a few quarters into the future as long as no major changes occur; and (ii) dynamic stochastic general equilibrium models (DSGE), which by construction assume a world always in equilibrium. The DSGE-models are actively used by central banks, which in part rely on them to take important decisions such as fixing interest rates. Both of these methods assume that the acting agents are fully rational and informed, and that their actions will lead to stable equilibria. These models therefore do not encompass out-of-equilibrium phenomena such as bubbles and subsequent crashes (Kindleberger, 2000; Sornette, 2003), arising among other mechanisms from herding among not fully rational traders (De Grauwe, 2010). Consequently, policy makers such as central banks base their expectations on models and processes that do not contain the full spectrum of possible outcomes and are caught off guard when extreme events such as the financial crisis in 2008 occur (Colander et al., 2009). Indeed, during and following the financial crisis of 2007-2008 in the US that cascaded to Europe in 2010 and to the world, central bankers in top policy making positions, such as Trichet, Bernanke, Turner and many others, have expressed significant dissatisfaction with economic theory in general and macroeconomic theory in particular, suggesting even that their irrelevance in times of crisis.

Physics as well as other natural sciences, in particular evolutionary biology and environmental sciences, 
may provide inspiring paths to break the stalemate. The analytical and computational
concepts and tools developed in physics in particular are starting to provide important 
frameworks for a revolution that is in the making. We refer in particular to the
computational framework using agent-based or computational economic models. In this respect, let us quote Jean-Claude Trichet, the previous chairman of the European Central Bank in 2010: ÒFirst, we have to think about how to characterize the homo economicus at the heart of any model. The atomistic, optimizing agents underlying existing models do not capture behavior during a crisis period. We need to deal better with heterogeneity across agents and the interaction among those heterogeneous agents. We need to entertain alternative motivations for economic choices. Behavioral economics draws on psychology to explain decisions made in crisis circumstances. Agent-based modeling dispenses with the optimization assumption and allows for more complex interactions between agents. Such approaches are worthy of our attention.Ó And, as Alan Kirman (2012) stressed recently, computational or algorithmic models have a long and distinguished tradition in economics.  The exciting result is that simple interactions at the micro level can generate sophisticated structure at the macro level, exactly as observed in financial time series. Moreover, such ABMs are not constrained to equilibrium conditions. Out-of-equilibrium states can naturally arise as a consequence of the agents' behavior, as well as fast changing external conditions and impacting shocks, and lead to dramatic regime shift or tipping points. The fact that such systemic phenomena can naturally arise in agent based models makes this approach ideal to model extreme events in financial markets. The emphasis on ABMs and computational economics parallels a similar 
revolution in Physics that developed over the last few decades. Nowadays, most physicists
would agree that Physics is based on three pillars: experiments, theory and numerical simulations, 
defining the three inter-related disciplines of experimental physics, theoretical physics
and computational physics. Many scientists have devoted their life in just one of these three.
In comparison, computational economics and agent-based models are still in their infancy
but with similar promising futures.

Given the above mentioned analogies and relationships between economics and physics,
it is noteworthy that these two fields have been life-long companions during their mutual
development of concepts and methods emerging in both fields. There has been
much mutual enrichment and catalysis of cross-fertilization.
Since the beginning of the formulation of the scientific approach in the
physical and natural sciences, economists have taken inspiration
from physics, in particular in its success in describing natural regularities
and processes.  Reciprocally, physics has been inspired several times by
observations done in economics. 

This review aims at providing some insights on this relationship, past, present and future.
In the next section, we present a selected
history of mutual fertilization between physics and economics.
Section 3 attempts to dissect the fundamentally different perspectives
embraced in theories developed in financial economics compared with physics.
For this, the excess volatility puzzle is presented and analyzed in some depth.
We explain the meaning of puzzles and the difference between
empirically founded science and normative science.
Section 4 reviews how the Ising model of phase transitions has
developed to model social and financial systems. In particular, 
we present the concept of random utilities and derive the logit model
describing decisions made by agents, 
as being the analog of the Boltzmann factor in statistical physics.
The Ising model in its simplest form can then be derived as the optimal
strategy for boundedly rational investors facing discrete choices. 
The section also summarises the recent developments on non-orthodox
decision theory, called quantum decision theory.
Armed with these concepts, section 5 reviews non-exhaustively
a wealth of models that build on the Ising model and generalize
it to account for the many stylized facts of financial markets, and more,
with still a rich future to enlarge the scope of the investigations.
Section 6 briefly reviews our work on financial bubbles and crashes
and how the Ising model comes into play. Section 7 covers
the literature on agent-based models, of which the class of Ising models
can be considered a sub-branch. This section also presents the
main challenges facing agent-based modelling before being widely 
adopted by economists and policy makers. We also formulate the
``Emerging Market Intelligence hypothesis'', to explain the pervasive
presence of ``noise traders'' together with the near efficiency
of financial markets. 
Section 8 concludes with advice on the need to combine concepts and tools 
beyond physics and finance with evolutionary biology.

\section{A short history of mutual fertilization between physics and economics}

Many physicists and economists have reflected on the relationships between physics and 
economists. Let us mention some prominent accounts (Zhang, 1999; Bouchaud, 2001;
Derman, 2004; Farmer and Lux, 2010). Here, we consider rather the 
history of the inter-fertilisation between the two fields, providing an hopefully general
inspiring perspective especially for the physicist aspiring to work in economics
and finance.

\subsection{From Isaac Newton to Adam Smith}

 To formulate his ``Inquiry into the Nature and Causes of the Wealth of Nations'', Adam Smith
(1776) was inspired by the Philosophiae Naturalis
 Principia Mathematica (1687) of Isaac Newton, which specifically stresses
the  (novel at the time) notion of causative forces. In the first half
of the nineteenth century, Quetelet  and Laplace among others become fascinated
by the regularities of social phenomena such as births, deaths, crimes and suicides,
even coining the term ``social physics'' to capture the evidence for
natural laws (such as the ubiquitous Gaussian distribution
based on the law of large numbers and the central limit theorem)
 that govern human social systems such as the economy.

\subsection{Equilibrium}

In the second half of the 19th century, the microeconomists Francis
 Edgeworth and Alfred Marshall drew on the concept of 
macroequilibrium in gas, understood to be the result of the multitude
of incessant micro-collisions of gas particles, which
was developed by Clerk Maxwell and Ludwig Boltzmann. 
Edgeworth and Marshall thus developed 
the notion that the economy achieves an equilibrium state
not unlike that described for gas. In the same way that 
the thermodynamic description of a gas at equilibrium produces a mean-field
homogeneous representation that gets rid of the rich heterogeneity of
the multitude of micro-states visited by all the particles, 
the dynamical stochastic general equilibrium (DSGE) models used by central banks
for instance do not have agent heterogeneity and focus on a representative
agent and a representative firm, in a way parallel to the
Maxwell Garnett effective medium theory of dielectrics and 
effective medium approximations for conductivity and wave propagation 
in heterogenous media. In DSGE, equilibrium refers to clearing markets, 
such that total consumption equal output, or total demand equals total supply,
and this takes place between representative agents.
This idea, which is now at the heart of economic modeling,
was not accepted easily by contemporary economists
who believed that the economic world is out-of-equilibrium, with
heterogeneous agents who learn and change their preferences
as a function of circumstances. It is important to emphasize
that the concept of equilibrium, which
has been much criticized in particular since the advent of the ``great
financial crisis'' since 2007 and of the ``great recession'', was the
result of a long maturation process with many fights within the economic profession.
In fact, the general equilibrium theory now at the core of mainstream economic
modeling is nothing but a formalization of the idea that ``everything in
 the economy affects everything else'' (Krugman, 1996), reminiscent of
 mean-field theory or self-consistent effective medium methods in
 physics. However, economics has pushed further than physics the role
of equilibrium by ascribing to it a normative role, i.e., not really
striving to describe economic systems as they are, but rather as
they should be (Farmer and Geanakoplos, 2009).

\subsection{Pareto and power laws}

 In his ``Cours d'Economie Politique''  (1897), the economist and philosopher 
Vilfredo Pareto reported remarkable regularities in 
the distribution of incomes, described by the eponym
 power laws, which have later become the focus of many natural scientists
and physicists attracted by the concept of universality and scale invariance (Stanley, 1999).
Going beyond Gaussian statistics, power laws belong to the class
of ``fat-tailed'' or sub-exponential distributions.

One of the most important implications of the existence of the fat-tail
nature of event size distributions is that the probability of observing
a very large event is not negligible, contrary to the prediction of 
the Gaussian world, which rules out for all practical purposes events 
with sizes larger than a few standard deviations from the mean. 
Fat-tailed distributions can even be such that the
variance and even the mean are not defined mathematically, corresponding
the wild class of distributions where the presence of extreme event sizes is intrinsic.

Such distributions have later been documented for many types of systems, 
when describing the relative frequency of the sizes of events they generate, 
for instance earthquakes, avalanches, landslides, storms, forest fires, solar flares, commercial
 sales, war sizes, and so on (Mandelbrot, 1982; Bak, 1996; Newman, 2005; Sornette, 2004).
Notwithstanding the appeal for a universal power law description, the reader
should be warned that many of the purported power law distributions are 
actually spurious or only valid over a rather limited range 
(see e.g. Sornette, 2004; Perline, 2005; Clauset et al., 2009).
Moreover, most data in finance show strong dependence, which invalidates 
simple statistical tests such as the Kolmogorov Smirnov test (Clauset et al., 2009).
A drastically different view point is offered by multifractal processes, such
as the multifractal random walk  (Bacry et al., 2001; 2013; Muzy et al., 2001; 2006),
in which the multiscale two-point correlation structure of the volatility is the primary 
construction brick, from which derives the power law property
of the one-point statistics, i.e. the distribution of returns (Muzy et al., 2006).
Moreover, the power law regime may even be superseded
by a different ``dragon-king'' regime in the extreme right tail (Sornette, 2009; Sornette and Ouillon, 2012).

\subsection{Brownian motion and random walks}

 In order to model the apparent random walk motion of bonds and stock options
 in the Paris stock market, mathematician Louis Bachelier (1900) developed in his thesis
 the mathematical theory of diffusion (and the first elements of
 financial option pricing). He solved the parabolic diffusion equation
 five years before Albert Einstein (1905) established the theory of
 Brownian motion based on the same diffusion equation,  also underpinning
 the theory of random walks. These two works have ushered research on
mathematical descriptions of fluctuation phenomena in 
statistical physics, of quantum fluctuation processes in elementary 
particles-fields physics, on the one hand, and of financial prices
on the other hand, both anchored in the random walk model
and Wiener process. The geometric Brownian motion (GBM) (exponential
of a standard random walk) was introduced by
Osborne (1959) on empirical grounds and Samuelson (1965)
on theoretical grounds that prices cannot become 
negative and price changes are proportional to previous prices.
Cootner (1964) compiled strong empirical support for the 
GBM model of prices and its associated
log-normal distribution of prices, corresponding to Gaussian
distributions of returns. The GBM model has become the backbone of financial economics theory,
underpinning many of its fundamental pillars, 
such as Markowitz' portfolio theory (Markowitz, 1952), Black-Scholes-Merton option 
pricing formula (Black and Scholes, 1973; Merton, 1973) and the Capital Asset Pricing Model 
(Sharpe, 1964) and its 
generalized factor models of asset valuations (Fama and French, 1993; Carhart, 1997). 
Similarly, it is not exaggerated to state that much of physics is occupied with modeling
fluctuations of (interacting) particles undergoing some kind of 
correlated random walk motion.
As in physics, empirical analyses of financial fluctuations have forced
the introduction of a number of deviations from the pure naive random walk model,
in the form of power law distribution of log-price increments, long-range
dependence of their absolute values (intermittency and clustering)
and absence of correlation of returns,
multifractality of the absolute value of returns
(multi-scale description due to the existence of information cascades) 
(Mandelbrot, 1997; Mandelbrot et al., 1997; Bacry et al., 2001)
and many others (Chakraborti et al., 2011). 
A profusion of models have been introduced to 
account for these observations, which build on the GBM model.

\subsection{Stable L\'evy distributions}

 In the early 1960s, mathematician Benoit Mandelbrot (1963) pioneered
 the use in Financial Economics of heavy-tailed distributions 
(stable L\'evy laws), which exhibit power law tails with exponent less than $2$\footnote{Heavy-tailed
distributions are defined in the mathematical literature (Embrechts et al., 1997)
roughly speaking by exhibiting a probability density function (pdf)
with a power law tail of the form pdf$(x) \sim 1/x^{1+\mu}$ 
with $0 < \mu < 2$ so that the variance and other
centered moments of higher orders do not exist.}, in contrast with
the traditional Gaussian (Normal) law.. Several
economists at the University of Chicago (Merton
 Miller, Eugene Fama, Richard Roll), at MIT (Paul Samuelson) and at
 Carnegie Mellon University (Thomas Sargent) were initially
attracted by Mandelbrot's suggestion to replace the Gaussian framework
by a new one based on stable L\'evy laws. 
In his PhD thesis, Eugene Fama confirmed that the frequency
 distribution of the changes in the logarithms of prices was
 ``leptokurtic'', i.e., with a high peak and fat tails. However, other
 notable economists (Paul Cootner and Clive Granger) strongly opposed
 Mandelbrot's proposal, based on the argument that ``the statistical
 theory that exists for the normal case is nonexistent for the other
 members of the class of LŽvy laws.''  Actually, Fama (1965), Samuelson (1967)
and later Bawa et al. (1979) extended Markowitz' portfolio theory to the case
of stable Paretian markets, showing that some of the standard concepts
and tools in financial economics have a natural generation
in the presence of power laws. This last statement has been made
firmer even in the presence of non-stable power law tail distributions by Bouchaud et al. (1998).
However, the interest in stable L\'evy laws faded as 
empirical evidence mounted rapidly to show that the distributions of returns 
are becoming closer to the Gaussian law at time scales larger than one month, 
in contradiction with the self-similarity hypothesis associated with the L\'evy laws
(Campbell et al., 1997; MacKenzie, 2006). 
In the late 1960s, Benoit Mandelbrot mostly stopped his research
in the field of financial economics. However, inspired by his forays
on the application of power laws to empirical data, 
he went on to show that non-differentiable geometries (that he coined
 ``fractal''), previously developed by mathematicians (Weierstrass, H\"older, 
Hausdorff among others) from the 1870s to
 the 1940s, could provide new ways to deal with the real complexity of
 the world (Mandelbrot, 1982).  This provided an inspiration for the
econophysicists' enthusiasm starting in the 1990s to model the
 multifractal properties associated with the long-memory properties
 observed in financial asset returns (Mandelbrot et al., 1997;
 Mandelbrot, 1997; Bacry et al., 2001; 2013; Muzy et al., 2001; 2006; Sornette et
 al., 2003).

\subsection{Power laws after Mandelbrot}

Much of the efforts in the econophysics literature of the late 1990s and early 2000s revisited and
refined the initial 1963 Mandelbrot hypothesis on
heavy-tailed distribution of returns, confirming on the one hand the existence of the
variance (which rules out the class of L\'evy distributions proposed by
Mandelbrot), but also suggesting a power law tail with an exponent $\mu$
close to $3$ (Mantegna and Stanley, 1995; Gopikrishnan et al., 1999).
Note however that several other groups have discussed alternatives, such as exponential
(Silva et al. (2004) or stretched exponential distributions (Laherrere
 and Sornette, 1999). Moreover, Malevergne et al. (2005) and  Malevergne and Sornette
 (2006; Chapter 2) developed an asymptotic statistical theory showing that the power law distribution
 is asymptotically nested within the larger family of stretched exponential distributions,
 allowing the use of the Wilks log-likelihood ratio statistics of nested hypotheses
 in order to decide between power law and stretched exponential for a given data set.
 Similarly, Malevergne et al. (2011) developed a uniformly most powerful unbiased
 test to distinguish between the power law and log-normal distributions, whose
 statistics turns out to be simply the sample coefficient of variation (the ratio
of the sample standard deviation (std) to the sample mean of the logarithm of the random
 variable).
 
Financial engineers actually care about these
 technicalities because the tail structure controls the Value-at-Risk
 and other risk measures used by regulators as well as investors
to assess the soundness of firms as well as the quality of investments.
Physicists care because the
 tail may constrain the underlying mechanism(s). For instance, Gabaix et
 al. (2003) attribute the large movements in stock market activity to
 the interplay between the power-law distribution of the sizes of large
 financial institutions and the optimal trading of such large
 institutions. Levy and Levy (2003) and Levy (2005) similarly emphasize
the importance of the Pareto wealth distribution in explaining the
distribution of stock returns, pointing out that the Pareto wealth distribution, 
market efficiency, and the power law distribution of stock returns are closely linked
and probably associated with stochastic multiplicative processes
(Sornette and Cont, 1997; Sornette, 1998a; Malevergne and Sornette, 2001;
Huang and Solomon, 2002;  Solomon and Richmond, 2002; Malcai et al., 2002; 
Lux and Sornette, 2002; Saichev et al., 2010).
However, another strand of literature emphasizes that most 
large events happen at relatively high frequencies, and 
seem to be triggered by a sudden drop in liquidity rather than to an outsized order
(Farmer et al., 2004; Weber and Rosenow, 2006; Gillemot et al., 2007; Joulin et al., 2008).

\subsection{Full distribution, positive feedbacks, inductive reasoning}

In a seminal Nobel prize winning article, Anderson (1958) laid out 
the foundation of the physics of heterogenous complex systems 
by stressing the need to go beyond the standard description in terms of the first two
moments (mean and variance) of statistical distributions. He pointed out the 
importance of studying their full shape in order to account for important rare large deviations
that often control the long-term dynamics and organization of
complex systems (dirty magnetic systems, spin-glasses). In the same vein, 
Gould (1996) has popularized the need to look 
at the ``full house'', the full distribution, in order to explain many paradoxes 
in athletic records as well as in the biology of evolution.
The study of spinglasses (M\'ezard et al., 1987) and of out-of-equilibrium 
self-organizing complex systems (Strogatz, 2003; Sornette, 2004; Sethna, 2006)
have started to inspire economists to break the stalemate associated
with the concept of equilibrium, with emphasis
on positive feedbacks and increasing returns (Arthur, 1994a; 1997; 2005; Krugman, 1996)
and on inductive bottom-up organizational processes (Arthur, 1994b; Challet et al., 2005).
This is in contrast with the deductive  top-down reasoning most often used in economics, leading
to the so-called ``normative'' approach of economics, which aims at providing recipes on how 
economies should be, rather than striving to describe how they actually are.

\section{Thinking as an economist or as a physicist?}

\subsection{Puzzles and normative science}

Economic modeling (and financial economics is just a branch following
the same principles) is based on the hunt for paradoxes or puzzles.
The term puzzle refers to  problems posed by empirical observations that do not conform to the 
predictions based on theory. Many puzzles have been unearthed by financial economists.
One of the most famous of these paradoxes
is called the excess volatility puzzle, which was discovered by Shiller (1981;1989)
and LeRoy and Porter (1981).

A puzzle emerges typically by the following procedure. 
A financial modeler builds a model or a class of models based on
a pillar of standard economic thinking, such as efficient markets,
rational expectations, representative agents, and so on. She then draws some 
prediction that is then tested statistically, often using linear regressions on empirical data.
A puzzle emerges when there is a strong divergence or disagreement between
the model prediction and the regressions, so that something seems at odds, 
literally ``puzzling'' when viewed from the interpreting lenses of the theory.
But rather than rejecting the model as the falsification process in physics
dictates (Dyson, 1988), the financial modeler is excited because she has hereby identified
a new ``puzzle'': the puzzle is that the ``horrible'' reality (to quote Huxley) does
not conform to the beautiful and parsimonious (and normative) theoretical edifice of 
neo-classical economic thinking. This is a puzzle because the theory should not
be rejected, it cannot be rejected, and therefore the data has something wrong in it, 
or there are some hidden effects that have to be taken into account that 
will allow the facts to confirm the theory when properly treated. 
In the most generous acceptation, a puzzle points to improvements to bring
to the theory. But the remarkable thing remains that the theory is not falsified.
It is used as the deforming lens to view and interpret empirical facts.

This rather critical account should be balanced with the benefits obtained
from studying ``puzzles'' in economics. Indeed, since it has the goal of
formalising the behavior of individuals and of organisations striving 
to achieve desired goals in the presence of scarce resources, economics
has played and is still playing a key role in helping policy makers 
shape their decision when governing organisation and nations.
To be concerned with how things should be may be a good idea, especially
with the goal of designing ``better'' systems. 
If and when reality deviates from the ideal, this signals to economists 
the existence of some ``friction'' that
needs to be considered and possibly alleviated. Frictions are important within
economics and, in fact, are often modelled.

\subsection{The volatility smile}

This ideology is no better illustrated than by the concept of the ``volatility smile''.
The celebrated Black-Scholes-Merton pricing formula calculates the value of options,
derivatives defined on underlying assets, such as the European call option that gives the right
but not the obligation for its holder to buy the underlying stock at some fixed exercise price $K$
at a fixed maturity time $T$ in the future (Black and Scholes, 1973; Merton, 1973). 
In addition to the exercise price $K$ and the time $T-t$ to maturity
counted from the present time $t$,  the Black-Scholes-Merton pricing formula depends on 
several other parameters, namely the risk-free interest rate, the volatility $\sigma$ of the 
returns of the underlying asset as well as its present price $p(t)$. 

As recounted by
MacKenzie (2006), the spreading use of the Black-Scholes-Merton pricing formula 
associated with the opening of the Chicago Board Options Exchange in 1973 led to
a progressive convergence of traded option prices to their Black-Scholes theoretical valuation,
legitimizing and catalyzing the booming derivative markets. This developed nicely until the crash of
19 October 1987, which, in one stroke, broke for ever the validity of the formula.
Since that day, one literally fudges the Black-Scholes-Merton 
formula by adjusting  the volatility parameter 
to a value $\sigma_{\rm implied}$ such that the Black-Scholes-Merton formula coincides with the
empirical price. The corresponding volatility value is called ``implied'', because it
is the value of $\sigma$ needed in the formula, and thus ``implied'' by the markets, in order
for theory and empirics to agree. The volatility smile refers to the fact that  $\sigma_{\rm implied}$ 
is not a single number, not even a curve, but rather a generally convex surface, function of both $K$ and $T-t$:
in order to reconcile the failing formula, one needs fudged values of $\sigma$
for all possible pairs of $K$ and $T-t$ traded on the market for each underlying asset.

This is contrast to the theory that assumes a single unique fixed value representing the 
standard deviation of the returns of the underlying asset.
The standard financial rationale is that the volatility smile $\sigma_{\rm implied}(K, T-t)$ quantifies
the aggregate market view on risks. Rather than improving the theory, the failed formula
is seen as the engine for introducing an effective risk metric that gauges the market risk perception
and appetites. Moreover, the volatility smile surface $\sigma_{\rm implied}(K, T-t)$
depends on time, which is interpreted as reflecting the
change of risk perceptions as a function of economic and market conditions.
This is strikingly different from the physical approach, which 
would strive to improve or even cure the  Black-Scholes-Merton failure
(Bouchaud and Sornette, 1994; Bouchaud and Potters, 2003) by accounting
for non-Gaussian features of the distribution of returns, long-range
dependence in the volatility as well as other market imperfections
that are neglected in the standard Black-Scholes-Merton theory.

The implied volatility type of thinking is so much ingrained that all
traders and investors are trained in this way, think according to the risks
supposedly revealed by the implied volatility surface and develop
correspondingly their intuition and operational implementations.
By their behaviors, the traders actually justify the present use of the implied volatility
surface since, in finance, if everybody believes in something, it will happen
by their collective actions, called self-fulfilling prophecies. 
It is this behavioral boundedly rational feedback of traders' perception on risk taking and hedging that
is neglected in the Black-Scholes-Merton theory. Actually, Potters et al. (1998)
showed, by studying in detail the market prices of options on liquid markets, 
that the market has empirically corrected the simple, but inadequate Black-Scholes formula 
to account for the fat tails and the correlations in the scale of fluctuations. 
These aspects, although not included in the pricing models, are found very precisely 
reflected in the price fixed by the market as a whole. 

Sircar and Papanicolaou (1998)
showed that a partial account of this feedback of hedging in the Black-Scholes theory
leads to increased volatility. Wyart and Bouchaud (2007) formulated a nice
simple model for self-referential behavior in financial markets where agents build strategies
based on their belief of the existence of correlation between some flow of information and prices.
Their belief followed by action makes the former realized and may produce excess
volatility and regime shifts that can be associated with the concept of convention (Orl\'ean, 1995).

\subsection{The excess volatility puzzle: thinking as an economist}
  
As another illustration of the fundamental difference between how economists and physicists
construct models and analyze empirical data, let us dwell further on the ``excess volatility puzzle'' 
discovered by Shiller (1981;1989) and LeRoy and Porter (1981), according to which
observed prices fluctuate much too much compared with what is expected from their
fundamental valuation.

Physics uses the concept
of causality: prices should derive from fundamentals. Thus,
let us construct our best estimate for the fundamental price $p^*(t)$.
The price, which should be a ``consequence'' of the 
fundamentals, should be an approximation of it. The physical idea
is that the dynamics of agents in their expectations and trading
should tend to get the right answer, that is, $p(t)$ should be
an approximation of $p^*(t)$. Thus, we write 
\be
p(t) = p^*(t) + \epsilon'(t)~,
\label{jghklssd}
\ee
and there is no excess volatility paradox. The large volatility of 
$p(t)$ compared with $p^*(t)$ provides an information on the price
forming processes, and in particular tells us that the dynamics of
price formation is not optimal from a fundamental valuation perspective.
The corollary is that prices move for other reasons than fundamental
valuations and this opens the door to investigating 
the mechanisms underlying price fluctuations.

In contrast, when thinking in equilibrium,
the notion of causality or causation ceases to a large degree to play a role in finance. According
to finance, it is not because the price should be the logical consequence
of the fundamentals that it should derive from it. In contrast, the
requirement of ``rational expectations''  (namely that
agents' expectations equal true statistical expected values) gives
a disproportionate faith in the market mechanism and collective agent behavior
so that, by a process similar to Adam Smith's invisible hand, 
the collective of agents by the sum of their actions, similar to the action
of a central limit theorem given an average converging with absolute
certainty to the mean with no fluctuation in the large $N$ limit, converge
to the right fundamental price with almost certainty. Thus, the observed
price is the right price and the fundamental price is only approximately 
estimated because not all fundamentals are known with good precision. And
here comes the excess volatility puzzle.

In order to understand all the fuss made in the name of the
excess volatility puzzle, we need to go back to the definition of value.
According the efficient market hypothesis (Fama, 1970; 1991; Samuelson, 1965; 1973),
the observed price $p(t)$ of a share (or of a
portfolio of shares representing an index) equals the mathematical expectation, conditional on all
information available at the time, of the present value $p^*(t)$ of actual subsequent dividends
accruing to that share (or portfolio of shares). This fundamental value $p^*(t)$ is not known at time $t$, and has to be
forecasted. The key point is that the efficient market hypothesis holds that 
the observed price equals the optimal forecast of it.
Different forms of the efficient markets model differ for instance in their choice of the discount rate 
used in the present value, but the general efficient markets model can be written as
\be
p(t)  = {\rm E}_t[p^*(t)]~,
\label{heyth}
\ee
where ${\rm E}_t$ refers to the mathematical expectation conditional 
on public information available at time $t$.
This equation asserts that any surprising movement in the stock market must have at its origin
some new information about the fundamental value $p^*(t)$.
It follows from the efficient markets model that 
\be
p^*(t) = p(t) + \epsilon(t)
\label{jghkld}
\ee
where $\epsilon(t)$ is a forecast error.
The forecast error $\epsilon(t)$ must be uncorrelated 
with any information variable available at time $t$,
otherwise the forecast would not be optimal; it would not be taking into account all information.
Since the price $p(t)$ itself constitutes a piece of information at time $t$, $p(t)$ and $\epsilon(t)$
must be uncorrelated with each other.
Since the variance of the sum of two uncorrelated variables is the sum of their variances, it
follows that the variance of $p^*(t)$ 
must equal the variance of $p(t)$ plus the variance of $\epsilon(t)$. Hence,
since the variance of $\epsilon(t)$ cannot be negative, one obtains 
that the variance of $p^*(t)$ must be greater than or equal
to that of $p(t)$. This expresses the fundamental principle of optimal forecasting,
according to which the forecast must be less
variable than the variable forecasted. 

Empirically, one observes that the volatility of the realized price 
$p(t)$ is much larger than the volatility of the fundamental price
$p^*(t)$, as estimated from all the sources of fluctuations of the 
variables entering in the definition of $p^*(t)$. This is the 
opposite of the prediction resulting from expression (\ref{jghkld}).
This disagreement between theoretical prediction and empirical
observation is then referred to as the ``excess volatility puzzle''.
This puzzle is considered by many financial economists as perhaps
the most important challenge to the orthodoxy of efficient markets
of neo-classical economics and many researchers have written on 
its supposed resolution. 
 
To a physicist, this puzzle is essentially non-existent. Rather than (\ref{jghkld}),
a physicist would indeed have written expression (\ref{jghklssd}),
that is, the observed price is an approximation of the fundamental price, up 
to an error of appreciation of the market. The difference between (\ref{jghkld})
and (\ref{jghklssd}) is at the core of the difference in the modeling strategies
of economists, that can be called top-down (or from rational expectations and 
efficient markets), compared with the bottom-up or microscopic approach of physicists.
According to equation (\ref{jghklssd}), the fact that the volatility of $p(t)$ is
larger than that of the fundamental price $p^*(t)$ is not a problem; it simply 
expresses the existence of a large noise component in the pricing mechanism.

Black (1985) himself introduced the notion of ``noise traders'', embodying the
presence of traders who are less than fully rational and whose influence
can cause prices and risk levels to diverge from expected levels. 
Models built on the analogy with the Ising model to capture
social influences between investors are reviewed in the next section, which often
provide explanations for the excess volatility puzzle. Let us mention 
in particular our own candidate in terms of the ``noise-induced volatility'' phenomenon
(Harras et al., 2012).

\section{The Ising model and financial economics}

\subsection{Roots and sources}

The Ising model, introduced initially as a  mathematical model of ferromagnetism
 in statistical mechanics (Brush, 1967), is now part of the common culture of physics, 
as the simplest representation of interacting elements with a finite number of
possible states. The model consists
of a large number of magnetic moments (or spins) connected 
by links within a graph, network or grid. In the simplest version, the spins can only
take two values ($\pm 1$), which represent the direction in which they point (up
or down). Each spin interacts with its direct neighbors,
tending to align together in a common direction, while the temperature 
tends to make the spin orientations random. Due to the fight between
the ordering alignment interaction and the disordering temperature,
the Ising model exhibits a non-trivial
phase transition in systems at and above two dimensions. Beyond ferromagnetism,
it has developed into different generalized forms that find interesting 
applications in the physics of ill-condensed matter such as spin-glasses (Mezard et al., 1987) and 
in neurobiology (Hopfield, 1982).

There is also a long tradition of using the Ising model and
its extensions to represent social interactions and organization
(Wiedlich, 1971; 1991; 2000; Callen and Shapero, 1974; Montroll and Badger, 1974; 
Galam et al., 1982; Orlean, 1995). Indeed, the
analogy between magnetic polarization and opinion polarization was presented in the
early 1970s by  Weidlich (1971), in the framework of ``Sociodynamics'', and later by Galam 
et al. (1982) in a manifesto for ``Sociophysics''. In this decade, several efforts
towards a quantitative sociology developed (Schelling, 1971; 1978; Granovetter, 1978; 1983),
based on models essentially undistinguishable from spin models.

A large set of economic models can be mapped onto various versions of
the Ising model to account for social influence in individual
decisions (see Phan et al. (2004) and references
therein). The Ising  model is indeed
one of the simplest models describing the competition between the
ordering force of imitation or contagion and the disordering impact
of private information or idiosyncratic noise, which leads already
to the crucial concept of spontaneous symmetry breaking and phase
transitions (McCoy and Wu, 1973). It is therefore not surprising to see
it appearing in one guise or another in models of social imitation (Galam and Moscovici, 1991)
and of opinion polarization (Galam, 2004; Sousa et al., 2005; Stauffer, 2005; Weidlich and Huebner, 2008).

The dynamical updating rules of
the Ising model can be shown to describe the formation of decisions
of boundedly rational agents (Roehner and Sornette, 2000) or
to result from optimizing agents whose utilities incorporate a
social component (Phan et al., 2004). 

An illuminating way to justify the use in social systems of the Ising model
(and of its many generalizations) together with a statistical physics approach
(in terms of the Boltzmann factor) derives from discrete choice models.
Discrete choice models consider as elementary entities the decision makers
who have to select one choice among a set of alternatives (Train, 2003). For instance, the choice
can be to vote for one of the candidates, or to find the right mate, or to attend a university among several
or to buy or sell a given financial asset. To develop the formalism of discrete 
choice models, the concept of a random utility is introduced, which is used to derive
the most prominent discrete choice model, the Logit model, which has 
a strong resemblance with Boltzmann statistics. The formulation of 
a binary choice model of socially interacting agents then allows one to obtain
exactly an Ising model, which establishes
a connection between studies on Ising-like systems in physics and collective behavior of
social decision makers.

\subsection{Random utilities, Logit model and Boltzmann factor}

In this section, our goal is to demonstrate the intimate link between 
the economic approach of random utilities and the framework of
statistical physics, on which the treatment of the Ising model in particular relies.

Random Utility Models provide a standard framework for
discrete choice scenarios. The decision maker has to choose one alternative
out of a set $X$ of $N$ possible ones. For each alternative $x \in X$, the decision maker
obtains the utility (or payoff) $U(x)$. The decision maker will choose the alternative
that maximizes his/her utility. However, neither an external observer nor the
decision maker herself may be fully cognizant of the exact form of the 
utility function $U(x)$. Indeed, $U(x)$ may depend upon a number of attributes
and explanatory variables, the environment as well as emotions, which 
are impossible to specify or measure exhaustively and precisely.
This is captured by writing
\begin{equation}
U(x) = V(x) + \epsilon(x)~,
\end{equation}
where $\epsilon(x)$ is the unknown part decorating the normative
utility $V(x)$. One interpretation is that $\epsilon(x)$ can represent the component of the
utility of a decision maker that is unknown or hidden to an observer trying
to rationalize the choices made by the decision maker, as done in 
experiments interpreted within the utility framework. Or $\epsilon(x)$
could also contain an intrinsic random part of the decision unknown
to the decision maker herself, rooted in her unconscious. 
As $\epsilon(x)$ is unknown to the researcher, it will be
assumed random, hence the name, random utility model.

The probability for the decision maker to choose $x$ over all other alternatives
$Y = X - \{x\}$ is then given by
\begin{eqnarray}
P(x) &=& {\rm Prob} \left( U(x) > U(y)~, ~ \forall y  \in Y \right) \nonumber \\
        &=& {\rm Prob} \left( V(x) - V(y)  > \epsilon(y) - \epsilon(x)    ~, ~ \forall y  \in Y \right)~.
\label{gth3yhj3y}
\end{eqnarray}
Holman and Marley (as cited in Luce and Suppes  (1965)) 
showed that if the unknown utility $\epsilon(x)$ is distributed according
to the double exponential distribution, also called the Gumbel distribution,
which has a cumulative distribution function (CDF) given by
\begin{equation}
F_G(x) = e^{-e^{-(x-\mu)/\gamma}}
\label{sthyjuju4}
\end{equation}
with positive constants $\mu$ and $\gamma$, then $P(x)$ defined in 
expression (\ref{gth3yhj3y}) is given by the logistic model, which
obeys the axiom of independence from irrelevant alternatives (Luce, 1959).
This axiom, at the core of standard utility theory, states that
the probability of choosing one possibility against another
from a set of alternatives is not affected by the addition or removal of other
alternatives, leading to the name ``independence from irrelevant alternatives''.

Mathematically, it can be expressed as follows.
Suppose that $X$ represents the complete set 
of possible choices and consider $S \subset X$, a subset of these choices. 
If, for any element $x \in X$, there is a finite probability 
$p_X(x) \in ]0; 1[$ of being chosen, then Luce's choice axiom is defined as
\begin{equation}
p_X(x) = p_S(x) \cdot p_X(S)~,
\label{etjyoo}
\end{equation}
where $p_X(S)$ is the probability of choosing any element in $S$ from the set $X$.
Writing expression (\ref{etjyoo}) for another element $y \in X$
and taking the ratios term by term leads to
\begin{equation}
{p_S(x) \over p_S(y)} = {p_X(x) \over p_X(y)}~,
\end{equation}
which is the mathematical expression of the axiom of
independence from irrelevant alternatives.
The other direction was proven by McFadden (1974), who showed that, if
the probability satisfies the independence from irrelevant alternatives condition, 
then the unknown utility $\epsilon(x)$ has to be
distributed according to the Gumbel distribution.

The derivation of the Logit model from expressions (\ref{gth3yhj3y})
and (\ref{sthyjuju4}) is as follows. In equation (\ref{gth3yhj3y}),
$P(x)$ is written
\begin{eqnarray}
P(x) &=& {\rm Prob} \left( V(x) - V(y) +\epsilon(x)  > \epsilon(y)    ~, ~ \forall y  \in Y \right)~,  \nonumber \\
        &=& \int_{-\infty}^{+\infty} \left( \prod_{y \in Y}   e^{-e^{-(V(x) - V(y) +\epsilon(x))/\gamma}} \right)
         f_G(\epsilon(x)) d\epsilon(x)~,
\label{gth3yhjwhy3y}
\end{eqnarray}
where $\mu$ has been set to $0$ with no loss of generality and 
$ f_G(\epsilon(x)) = {1 \over \gamma} e^{-x/\gamma}  e^{-e^{-x/\gamma}}$ is the probability
density function (PDF) associated with the CDF (\ref{sthyjuju4}).
Performing the change of variable $u = e^{-\epsilon(x)/\gamma}$, we have
\begin{eqnarray}
P(x) &=& \int_{0}^{+\infty} \left( \prod_{y \in Y}   e^{-u e^{ (V(x) - V(y))/\gamma}} \right) e^{-u} du~, \nonumber \\
	&=& \int_{0}^{+\infty}   e^{-u \sum_{y \in Y}  e^{- (V(x) - V(y))/\gamma}}  e^{-u} du~, \nonumber \\
	&=& {1 \over 1 + e^{- V(x)/\gamma} \sum_{y \in Y}  e^{V(y)/\gamma}} ~.
\label{gth3yhjqtw4hy3y}
\end{eqnarray}
Multiplying both numerator and denominator of the last expression (\ref{gth3yhjqtw4hy3y})
by $e^{V(x)/\gamma}$, keeping in mind that $Y = X - x$, the well known logit formulation is recovered,
\be
P(x) =  {e^{V(x)/\gamma} \over  \sum_{y \in X}  e^{V(y)/\gamma}} ~,
\label{rhtjuykjyue}
\ee
which fulfills the condition of independence from irrelevant alternatives.
Note that the Logit probability (\ref{rhtjuykjyue}) has the same form as 
the Boltzmann probability in statistical physics for a system to be found in a state of 
with energy $-V(x)$ at a given temperature $\gamma$.

\subsection{Quantum decision theory}

There is a growing realisation that even these above frameworks do not 
account for the many fallacies and paradoxes plaguing standard
decision theories (see for instance \url{http://en.wikipedia.org/wiki/List_of_fallacies}).
A strand of literature has been developing since about 2006 that borrows the
concept of interference and entanglement used in quantum mechanics
in order to attempt to account for theses paradoxes (Busemeyer et al., 2006; Pothos and Busemeyer, 2009).
A recent monograph reviews the developments using simple 
analogs of standard physical toy models, such as the two entangled spins
underlying the Einstein-Poldovsky-Rosen phenomenon (Busemeyer and Bruza, 2012).

From our point of view, the problem however is that these proposed remedies
are always designed for the specific fallacy or paradox under consideration and require a specific
set-up that cannot be generalised. To address this, Yukalov and Sornette (2008-2013)
have proposed a general framework, which extends
the interpretation of an intrinsic random component in any decision by 
stressing the importance of formulating the problem in terms of composite prospects.
The corresponding ``quantum decision theory" (QDT) is
based on the mathematical theory of separable Hilbert spaces. 
We are not suggesting that the brain operates according to the rule
of quantum physics. It is just that the mathematics of Hilbert spaces,
used to formalized quantum mechanics, provides the simplest generalization
of the probability theory axiomatized by Kolmogorov, which allows
for entanglement.
This mathematical structure captures the effect of superposition 
of composite prospects, including many incorporated intentions, 
which allows one to describe a variety of interesting fallacies 
and anomalies that have been reported to particularize the decision 
making of real human beings. The theory characterizes 
entangled decision making, non-commutativity of subsequent decisions, 
and intention interference. 

Two ideas form the basement of the QDT developed by Yukalov and Sornette (2008-2013).
First, our decision may be intrinsically probabilistic, i.e., when confronted
with the same set of choices (and having forgotten), we may choose different
alternatives. Second, the attraction to a given option (say choosing where to vacation
among the following locations: Paris, New York, Roma, Hawaii or Corsica) will depend in significant
part on the presentation of the other options, reflecting a genuine ``entanglement''
of the propositions. The consideration of composite prospects using
the mathematical theory of separable Hilbert spaces provides a natural
and general foundation to capture these effects. 
Yukalov and Sornette (2008-2013) demonstrated how the violation of the 
Savage's sure-thing principle (disjunction effect) can be explained 
quantitatively as a result of the interference of intentions, 
when making decisions under uncertainty. The sign and amplitude of 
the disjunction effects in experiments are accurately predicted using 
a theorem on interference alternation, which connects 
aversion-to-uncertainty to the appearance of negative interference 
terms suppressing the probability of actions. The conjunction fallacy 
is also explained by the presence of the interference terms. A series 
of experiments have been analysed and shown to be in excellent agreement with 
a priori evaluation of interference effects. The conjunction fallacy 
was also shown to be a sufficient condition for the disjunction effect 
and novel experiments testing the combined interplay between the two 
effects are suggested.

Our approach is based on the von Neumann theory of quantum measurements
(von Neumann, 1955), but with an essential difference. In quantum theory,
measurements are done over {\it passive} systems, while in decision theory,
decisions are taken by {\it active} human beings. Each of the latter is
characterized by its own {\it strategic state of mind}, specific for the
given decision maker. Therefore, expectation values in quantum decision theory 
are defined with respect to the decision-maker strategic state. In contrast, in
standard measurement theory, expectation values are defined through an arbitrary
orthonormal basis.

In order to give a feeling of how QDT works in practice, let us delineate 
its scheme. We refer to the published papers (Yukalov and Sornette, 2008, 
2009a,b,c; 2010a,b; 2011) for more in-depth presentations and preliminary tests. 
The first key idea of QDT is to consider the so-called prospects, which are the 
targets of the decision maker. Let a set of prospects $\pi_j$ be given, 
pertaining to a complete transitive lattice 
\be
\label{3}
\cL \equiv \{ \pi_j : \; j = 1,2,\ldots, N_L \} \;  .
\ee 
The aim of decision making is to find out which of the prospects is the 
most favorable. 

There can exist two types of setups. One is when a number of agents, say $N$, 
choose between the given prospects. Another type is when a single decision 
maker takes decisions in a repetitive manner, for instance taking decisions $N$ 
times. These two cases are treated similarly.

To each prospect $\pi_j$, we put into correspondence a vector $|\pi_j>$ in the 
Hilbert space, $\mathcal M$, called the mind space, and the prospect operator
$$
\hat P(\pi_j) \equiv | \; \pi_j \; \rgl \lgl \; \pi_j \; | \;   .
$$
QDT is a probabilistic theory, with the prospect probability defined as the 
average
$$
p(\pi_j) \equiv \lgl \; s \; | \; \hat P(\pi_j) \; | \; s \; \rgl
$$
over the strategic state $|s>$ characterizing the decision maker. 

Though some intermediate steps of the theory may look a bit complicated, the 
final results are rather simple and can be straightforwardly used in practice.   
Thus, for the prospect probabilities, we get finally
\be
\label{4}
 p(\pi_j) = f(\pi_j) + q(\pi_j) \;  ,
\ee
whose set defines a probability measure on the prospect lattice $\mathcal{L}$, 
such that
\be
\label{5}
 \sum_{\pi_j \in \cL} \; p(\pi_j) = 1 \; , \qquad
0 \leq p(\pi_j) \leq 1 \; .
\ee
The most favorable prospect corresponds to the largest of probabilities (\ref{4}).

The first term on the right-hand side of Eq. (\ref{4}) is the utility factor defined as
\be
\label{6}
f(\pi_j) \equiv \frac{U(\pi_j)}{\sum_j U(\pi_j) }
\ee
through the expected utility $U(\pi_j)$ of prospects. The utilities are 
calculated in the standard way accepted in classical utility theory. By this 
definition
$$
\sum_{\pi_j \in\cL} f(\pi_j) = 1 \; , \qquad 0 \leq f(\pi_j) \leq 1 \;   .
$$

The second term is an attraction factor that is a contextual object describing
subconscious feelings, emotions, and biases, playing the role of hidden variables. 
Despite their contextuality, it is proved that the attraction factors always 
satisfy the {\it alternation property}, 
such that the sum
\be
\label{7}
\sum_{\pi_j \in \cL} \; q(\pi_j) = 0 \; \qquad (-1 \leq q(\pi_j) \leq 1)
\ee
over the prospect lattice $\mathcal{L}$ be zero. In addition, the average 
absolute value of the attraction factor is estimated by the {\it quarter law}
\be
\label{8}
\frac{1}{N_L} \; \sum_{\pi_j\in \cL} \;
|\; q(\pi_j) \; | = \frac{1}{4} \;   .
\ee
These properties (\ref{7}) and (\ref{8}) allow us to {\it quantitatively} define the 
prospect probabilities (\ref{4}).

The prospect $\pi_1$ is more useful than $\pi_2$, when $f(\pi_1) > f(\pi_2)$. 
And $\pi_1$ is more attractive than $\pi_2$ , if $q(\pi_1) > q(\pi_2)$. 
The comparison between the attractiveness of prospects is done on the basis of 
the following criteria: more certain gain, more uncertain loss, higher activity 
under certainty, and lower activity under uncertainty and risk.   

Finally, decision makers choose the preferable prospect, whose probability (\ref{4}) 
is the largest. Therefore, a prospect can be more useful, while being less 
attractive, as a result of which the choice can be in favor of the less useful 
prospect. For instance, the prospect $\pi_1$ is preferable over $\pi_2$ when
\be
\label{9}
 f(\pi_1) - f(\pi_2) > q(\pi_2) - q(\pi_1) \;  .
\ee
This inequality illustrates the situation and explains the appearance of 
paradoxes in classical decision making, while in QDT such paradoxes never arise.  

The existence of the attraction factor is due to the choice under risk and 
uncertainty. If the latter would be absent, we would return to the classical
decision theory, based on the maximization of expected utility. Then we would 
return to the variety of known paradoxes.  

The comparison with experimental data is done as follows. Let $N_j$ agents
of the total number $N$ choose the prospect $\pi_j$. Then the aggregate 
probability of this prospect is given (for a large number of agents) by the 
frequency 
\be
\label{10}
 p_{exp}(\pi_j) = \frac{N_j}{N} \;  .
\ee
This experimental probability is to be compared with the theoretical prospect
probability (\ref{4}), using the standard tools of statistical hypothesis testing.
In this way, QDT provides a practical scheme that can be applied to realistic
problems. The development of the scheme for its application to various kinds
of decision making in psychology, economics, and finance, including temporal 
effects, provides interesting challenges.

Recently, Yukalov and Sornette (2013a) 
have also been able to define quantum probabilities of composite events,
thus introducing for the first time a rigorous and coherent
generalisation of the probability of joint events.
This problem is actually of high importance for the theory of quantum measurements 
and for quantum decision theory that is a part of measurement theory. 
Yukalov and Sornette (2013a) showed that
Luders probability of consecutive measurements is a transition probability between 
two quantum states and that this probability cannot be treated as a 
quantum extension of the classical conditional probability. Similarly,
the Wigner distribution was shown to be a weighted transition probability 
that cannot be accepted as a quantum extension of the classical joint probability. 
Yukalov and Sornette (2013a) suggested the definition of quantum joint 
probabilities by introducing composite events in multichannel measurements. 
Based on the notion of measurements under uncertainty, 
they demonstrated that the necessary condition for mode interference is 
the entanglement of the composite prospect together with the entanglement 
of the composite statistical state. Examples of applications
include quantum games and systems with multimode states, such as atoms, molecules, quantum dots, 
or trapped Bose-condensed atoms with several coherent modes (Yukalov et al., 2013).

\subsection{Discrete choice with social interaction and Ising model}

Among the different variables that influence the utility of the decision maker,
partial information, cultural norms as well as herding tend to push her decision
towards that of her acquaintances as well as that of the majority.
Let us show here how access to partial information and rational 
optimization of her expected payoff leads to strategies described by the Ising model 
(Roehner and Sornette, 2000).

Consider $N$ traders in a social network, whose links represent the communication
channels between the traders. 
We denote $N(i)$ the number of traders directly connected to $i$ in the network.
The traders buy or sell one asset at price $p(t)$, which evolves as a function of
time assumed to be discrete with unit time step.
 In the simplest version of the model, each agent can either buy or sell only one unit of the asset. 
 This is quantified by the buy state $s_i=+1$ or the sell state $s_i=-1$. Each agent can trade at
time $t-1$ at the price $p(t-1)$ based on all previous information up to $t-1$. 
We assume that the asset price variation is determined by the following equation
\be
{p(t)-p(t-1) \over p(t-1)} =  F\left({\sum_{i=1}^N s_i(t-1) \over N}\right) ~+~
 \sigma~ \eta(t)~,  \label{jfjfjka}
\ee
where $\sigma$ is the price volatility per unit time and $\eta(t)$ is a white Gaussian noise with
unit variance that represents for instance the impact resulting from the flow of 
exogenous economic news. 

The first term in the r.h.s. of (\ref{jfjfjka}) is the price impact function
describing the possible imbalance between buyers and sellers. We assume that the function $F(x)$
is such that $F(0)=0$ and is monotonically increasing with its argument. 
Kyle (1985) derived his famous linear price impact function $F(x) =\lambda x$
within a dynamic model of a market with a single risk neutral insider, random noise traders,
and competitive risk neutral market makers with sequential auctions.
Huberman and Stanzl (2004) later showed that,
when the price impact of trades is permanent and time-independent, only linear price-impact functions 
rule out quasi-arbitrage, the availability of trades that generate infinite expected profits.
We note however that this normative linear price impact impact has been 
challenged by physicists. Farmer et al. (2013)
report empirically that market impact is a concave function of the size of large trading orders.
They rationalize this observation as resulting from the algorithmic execution of 
splitting large orders into small pieces and executing incrementally. 
The approximate square-root impact function has been earlier rationalized by Zhang (1999)
with the argument that the time needed to complete a trade of size $L$ is proportional to $L$ 
and that the unobservable price fluctuations obey a diffusion process during that time. 
Toth et al. (2011) propose that the concave market impact function 
reflects the fact that markets operate in a critical regime
where liquidity vanishes at the current price, in the sense that all buy orders at price less
than current prices have been satisfied, and all sell orders at prices larger than the current
price have also been satisfied. The studies (Bouchaud et al., 2009; Bouchaud, 2010),
which distinguish between temporary and permanent 
components of market impact, show important links between impact function, the distribution of order sizes,
optimization of strategies and dynamical equilibrium. Kyle (private communication, 2012) and 
Gatheral and Schied (2013) point out that the issue is far from resolved
due to price manipulation, dark pools, predatory trading and no well-behaved optimal order
execution strategy. 
 
Returning to the implication of expression (\ref{jfjfjka}), at 
time $t-1$, just when the price $p(t-1)$ has been announced, the trader $i$
defines her strategy $s_i(t-1)$ that she will hold from $t-1$ to $t$, thus
realizing the profit $(p(t)-p(t-1))s_i(t-1)$. To define $s_i(t-1)$, the trader
calculates her expected profit ${\rm E}[P\&L]$, given the past information and her
position, and then chooses $s_i(t-1)$ such that ${\rm E}[P\&L]$ is maximal. Within the rational
expectation model, all traders have full knowledge of the fundamental equation
(\ref{jfjfjka}) of their financial world. However, they cannot poll the
positions $\{s_j\}$ that all other traders will take, which will determine the price drift
according to expression (\ref{jfjfjka}). The next best thing that trader $i$ can
do is to poll her $N(i)$ ``neighbors'' and construct her prediction for the
price drift from this information. The trader needs an additional
information, namely the a priori probability $P_+$ and $P_-$ for each trader to buy or sell.
The probabilities $P_+$ and $P_-$ are the only pieces of information that she can
use for all the traders that she does not poll directly. From this, she can form
her expectation of the price change. The simplest case corresponds to a
neutral market where $P_+=P_-=1/2$. To allow for a simple discussion, we restrict
the discussion to the linear impact function $F(x) = \lambda x$.  
The trader $i$ thus expects the following price change
\be
 \lambda~ \left({\sum_{j=1}^{*~N(i)} s_j(t-1) \over N}\right) +  \sigma~ {\hat \eta}_i(t)~,
\ee
where the index $j$ runs over the neighborhood of agent $i$
and ${\hat \eta}_i(t)$ represents the idiosyncratic perception of the economic 
news as interpreted by agent $i$.
Notice that the sum is now restricted to the $N(i)$ neighbors of trader $i$
because the sum over all other traders, whom she cannot poll directly, averages
out. This restricted sum is represented by the star symbol. Her expected profit is thus
\be
{\rm E}[P\&L] = \left(\lambda~ \left({\sum_{j=1}^{*~N(i)} s_j(t-1) \over N}\right) +
 \sigma~ {\hat \eta}_i(t)\right)~p(t-1)~s_i(t-1)~.
\ee
The strategy that maximizes her profit is
\be
s_i(t-1) = {\rm sign} \left( {\lambda \over N} \sum_{j=1}^{N(i)_*} s_j(t-1) ~+~
\sigma ~ {\hat \eta}_i(t)\right)~. \label{jgfhha}
\ee
Equation (\ref{jgfhha}) is nothing but the kinetic Ising model with Glauber dynamics
if the random innovations ${\hat \eta}_i(t)$ are distributed with a Logistic distribution
(see demonstration in the Appendix of (Harras et al., 2012)). 

This evolution equation
(\ref{jgfhha}) belongs to the class of stochastic dynamical models of interacting particles
(Liggett, 1995; 1997), which have been much studied mathematically in the
context of physics and biology. In this model (\ref{jgfhha}), the tendency
towards imitation is governed by $\lambda/N$, which is called the coupling
strength; the tendency towards idiosyncratic behavior is governed by $\sigma$. Thus
the value of $\lambda/N$ relative to $\sigma$ determines the outcome of the battle
between order (imitation process) and disorder, and the development of collective
behavior.  More generally, expression (\ref{jgfhha}) provides a convenient
formulation to model imitation, contagion and herding and many generalizations 
have been studied that we now briefly review.

\section{Generalized kinetic Ising model for financial economics}

The previous section motivates the notion that the Ising model
provides a natural framework to study the collective behaviour 
of interacting agents. Many generalisations have been introduced
in the literature and we provide a brief survey here.

The existence of an underlying Ising phase
transition, together with the mechanism of ``sweeping of an instability'' 
(Sornette, 1994; Stauffer and Sornette, 1999; Sornette et al., 2002), was found to
lead to the emergence of collective imitation that translate into the formation
of transient bubbles, followed by crashes (Kaizoji et al., 2002).

Bouchaud and Cont (1998) presented a nonlinear Langevin equation of 
the dynamics of a stock price resulting from the unbalance between supply
and demand, themselves based on two opposite opinions (sell and buy).
By taking into account the feedback effects of price variations
onto themselves, they find a formulation analogous to an inertial particle
in a quartic potential as in the mean-field theory of phase transitions.
 
Brock and Durlauf (1999) constructed a stylized model of community theory choice
based on agents' utilities that contains a term quantifying the degree of 
homophily which, in a context of random utilities, lead to a formalism
essentially identical to the mean field theory of magnetism. They find that 
periods of extended disagreement alternate with periods of rapid consensus formation,
as a result of choices that are made based on comparisons between
pairs of alternatives.  Brock and Durlauf (2001) further extend their model
of aggregate behavioral outcomes, in the presence of individual utilities
that exhibits social interaction effects, to the case of 
generalized logistic models of individual choice that
incorporate terms reflecting the desire of individuals to conform to the behavior of others in an environment of non-cooperative decision making. A multiplicity of equilibria is found when the social interactions 
exceed a particular threshold and decision making is non-cooperative. As expected
from the neighborhood of phase changes, a large susceptibility translates into the observation that
small changes in private utility lead to large equilibrium changes in average behavior.
The originality of Brock and Durlauf (2001) is to be able to 
introduce heterogeneity and uncertainty into the microeconomic specification of decision making, 
as well as to derive an implementable likelihood function that allows one to 
calibrate the agent-based model onto empirical data.

Kaizoji (2000) used an infinite-range Ising model to embody the tendency of traders to be
influenced by the investment attitude of other traders, which gives rise to regimes of bubbles
and crashes interpreted as due to the collective behavior of the agents at the Ising phase
transition and in the ordered phase. Biased
agent's idiosyncratic preference corresponds to the existence of an effective ``magnetic field'' 
in the language of physics. Because the social interactions compete with the biased
preference, a first-order transition exists, which is associated with the existence of crashes. 

Bornholdt (2001) studied a simple spin model in which traders interact at different
scales with interactions that can be of opposite signs, thus leading to ``frustration'',
and traders are also related to each other via their aggregate impact on the price.
The frustration causes metastable dynamics with intermittency and phases of chaotic dynamics,
including phases reminiscent of financial bubbles and crashes. While the model exhibits
phase transitions, the dynamics deemed relevant to financial markets is sub-critical.
 
Krawiecki et al. (2002)
used an Ising model with stochastic coupling coefficients, which leads to
volatility clustering and a power law distribution of returns
at a single fixed time scale.

Michard and Bouchaud (2005) have used the
framework of the Random Field Ising Model, interpreted as a threshold model
for collective decisions accounting both for agent heterogeneity and social imitation,
to describe imitation and social pressure found in data from
three different sources: birth rates, sales of cell phones and the drop of applause in concert halls.
 
Nadal et al. (2005) developed a simple market model with binary choices and 
social influence (called ``positive externality'' in economics), where the heterogeneity 
is either of the type represented by the Ising model at finite
temperature (known as annealed disorder) in a uniform external field
(the random utility models of Thurstone) or is fixed and corresponds to a 
a particular case of the quenched disorder model known as a random field Ising model, at zero temperature
(called the McFadden and Manski model). A novel first-order transition between 
a high price and a small number of buyers, to another one
with a low price and a large number of buyers, arises when the social influence is strong enough.
Gordon et al. (2009) further extend this model 
to the case of socially interacting individuals that make a binary choice in a context of positive additive endogenous externalities. Specifically, the different possible equilibria depend on the
distribution of idiosyncratic preferences, called here Idiosyncratic Willingnesses to Pay (IWP)
and there are regimes where several equilibria coexist, associated with non-monotonous demand function
as a function of price. This model is again strongly reminiscent of the random field Ising model studied 
in the physics literature.

Grabowski and Kosinski (2006) modeled the process of opinion formation in the human
population on a scale-free network, taking into account a
hierarchical, two-level structures of interpersonal interactions, as well as a spatial
localization of individuals. With Ising-like interactions together with a coupling
with a mass media ``field'', they observed several transitions and limit cycles, 
with non-standard ``freezing of opinions by heating'' and the rebuilding of 
the opinions in the population by the influence of the mass media at large annealed
disorder levels (large temperature).

Sornette and Zhou (2006a) and Zhou and Sornette (2007) 
generalized a stochastic dynamical formulation of the Ising model  (Roehner and Sornette, 2000)
to account for the fact that the imitation strength between agents may evolve 
in time with a memory of how past news have explained realized market returns. 
By comparing two versions of the
model, which differ on how the agents interpret the predictive power of news, 
they show that the stylized facts of financial
markets are reproduced only when agents are overconfident and mis-attribute the success of news 
to predict return to the existence of herding effects, thereby providing 
positive feedbacks leading to the model functioning close to the critical point.
Other stylized facts, such as a multifractal structure characterized 
by a continuous spectrum of exponents of the power law relaxation of
endogenous bursts of volatility, are well reproduced by this model of adaptation 
and learning of the imitation strength.
Harras et al. (2012) examined a different version of the Sornette-Zhou (2006a) formulation
to study the influence of a rapidly varying external signal to the Ising collective dynamics
for intermediate noise levels. They discovered the phenomenon of 
``noise-induced volatility'', characterized by an increase of the level of fluctuations 
in the collective dynamics of bistable units in the presence of a rapidly varying external signal. 
Paradoxically, and different from ``stochastic resonance'', the response of the system
becomes uncorrelated with the external driving force. Noise-induced volatility 
was proposed to be a possible cause
of the excess volatility in financial markets, of enhanced effective temperatures in a variety of out-of-equilibrium
systems, and of strong selective responses of immune systems of complex biological organisms. 
Noise-induced volatility is robust to the existence of various network topologies.

Horvath and Kuscsik (2007) considered a network with reconnection dynamics, 
with nodes representing decision makers modeled as
(``intra-net'') neural spin network with local and global inputs and feedback connections.
The coupling between the spin dynamics and the network rewiring produces
several of the stylized facts of standard financial markets, including the Zipf law for wealth.

Biely et al. (2009) introduced an Ising model in which spins are dynamically coupled by links in a dynamical network
in order to represent agents who are free to choose their interaction partners. 
Assuming that agents (spins) strive to minimize an ``energy'', the spins 
as well as the adjacency matrix elements organize together, 
leading to an exactly soluble model with reduced complexity
compared with the standard fixed links Ising model.
  
Motivated by market dynamics, 
Vikram and Sinha (2011) extend the Ising model by assuming that 
the interaction dynamics between individual components is mediated by a global variable making
the mean-field description exact. 

Harras and Sornette (2011) studied
a simple agent-based model of bubbles and crashes to clarify how 
their proximate triggering factor relate to
their fundamental mechanism. Taking into account three sources of information, (i) public
information, i.e. news, (ii) information from their ``friendship'' network and (iii) private
information, the boundedly rational agents continuously adapt their trading strategy to the
current market regime by weighting each of these sources of information in their trading
decision according to its recent predicting performance. 
In this set-up, bubbles are found to originate from
a random lucky streak of positive news, which, due to a feedback mechanism of these news
on the agents' strategies develop into a transient collective herding regime. 
Paradoxically, it is the attempt for investors to adapt to the current market regime that
leads to a dramatic amplification of the price volatility. A positive feedback loop is created
by the two dominating mechanisms (adaptation and imitation), which, by reinforcing each
other, result in bubbles and crashes. The model offers a simple reconciliation of the two
opposite (herding versus fundamental) proposals for the origin of crashes within a single
framework and justifies the existence of two populations in the distribution of returns,
exemplifying the concept that crashes are qualitatively different from the rest of the price
moves (Johansen and Sornette, 1998; 2001/2002; Sornette, 2009; Sornette and Ouillon, 2012).

Inspired by the bankruptcy of Lehman Brothers and its consequences on the global financial
system, Sieczka et al. (2011) developed a simple model in which the credit rating grades
of banks in a network of interdependencies follow a kind of Ising dynamics 
of co-evolution with the credit ratings of the other firms.
The dynamics resembles the evolution of a Potts spin glass with the external global field corresponding to a panic
effect in the economy. They find a global phase transition, between paramagnetic and ferromagnetic
phases, which explains the large susceptibility of the system to negative shocks. 
This captures the impact of the Lehman default event, quantified as having an almost
immediate effect in worsening the credit worthiness of all financial institutions in the economic network.
The model is amenable to testing different policies. For instance, 
bailing out the first few defaulting firms does not solve the problem, but does have the effect of 
alleviating considerably the
global shock, as measured by the fraction of firms that are not defaulting as a consequence.

Kostanjcar and Jeren (2013) defined a generalized Ising model of financial markets
with a kind of minority-game payoff structure and strategies that depend on order sizes.
Because their agents focus on the change of their wealth, they find that the macroscopic
dynamics of the aggregated set of orders (reflected into the market returns) remains stochastic
even in the thermodynamic limit of a very large number of agents.

Bouchaud (2013) proposed a general strategy for modeling collective socio-economic phenomena
with the Random Field Ising model (RFIM) and variants, which is argued to
provide a unifying framework to account for the existence of sudden ruptures and crises. 
The variants of the RFIM capture destabilizing self-referential feedback
loops, induced either by herding or trending. An interesting insight is the
determination of conditions under which Adam
Smith's invisible hand can fail badly at solving simple coordination problems.
Moreover, Bouchaud (2013) stresses that most of these models
assume explicitly or implicitly the validity of the so-called
``detailed-balance'' in decision rules, which is not a priori necessary
to describe real decision-making processes. The question of the robustness
of the results obtained when detailed balance holds to models where it
does not remain largely open. Examples from physics suggest that
much richer behaviors can emerge.

Kaizoji et al. (2013) introduced a model of financial bubbles 
with two assets (risky and risk-free), in which rational investors and noise 
traders co-exist. Rational investors form expectations on the return and risk of a risky asset and 
maximize their expected utility with respect to their allocation
on the risky asset versus the risk-free asset. Noise traders are subjected to social
imitation (Ising like interactions) and follow momentum trading (leading to 
a kind of time-varying magnetic field). Allowing for random time-varying
herding propensity as in (Sornette, 1994; Stauffer and Sornette, 1999; Sornette et al., 2002), 
this model reproduces the most important stylized facts
of financial markets such as a fat-tail distribution of returns, volatility clustering
as well as transient faster-than-exponential bubble growth with 
approximate log-periodic behavior (Sornette, 1998b; 2003).
The model accounts well for the behavior of traders and for the price
dynamics that developed during the dotcom bubble in 1995-2000.
Momentum strategies are shown to be transiently profitable, supporting
these strategies as enhancing herding behavior.

\section{Ising-like imitation of noise traders and models of financial bubbles and crashes}

\subsection{Phenomenology of financial bubbles and crashes}

Stock market crashes are momentous financial events that are fascinating
to academics and practitioners alike. According to the standard academic
textbook world view that markets are efficient, only the revelation of a
dramatic piece of information can cause a crash, yet in reality even the
most thorough post-mortem analyses are, for most large losses, inconclusive as to what
this piece of information might have been. For traders and investors,
the fear of a crash is a perpetual source of stress, and the onset of
the event itself always ruins the lives of some of them. Most approaches
to explain crashes search for possible mechanisms or effects that
operate at very short time scales (hours, days or weeks at most). Other
researchers have suggested market crashes may have endogenous origins.

In a culmination of almost 20 years of research in financial economics,
we have challenged the standard economic view that
stock markets are both efficient and unpredictable. We propose that the main concepts
that are needed to understand stock markets are imitation, herding,
self-organized cooperativity and positive feedbacks, leading to the
development of endogenous instabilities. According to this theory, local
effects such as interest raises, new tax laws, new regulations and so
on, invoked as the cause of the burst of a given bubble leading to a
crash, are only one of the triggering factors but not the fundamental
cause of the bubble collapse. We propose that the true origin of a
bubble and of its collapse lies in the unsustainable pace of stock
market price growth based on self-reinforcing over-optimistic
anticipation. As a speculative bubble develops, it becomes more and more
unstable and very susceptible to any disturbance.

In a given financial bubble, it is the expectation of future earnings
rather than present economic reality that motivates the average
investor. History provides many examples of bubbles driven by unrealistic
expectations of future earnings followed by crashes. The same basic
ingredients are found repeatedly. Markets go through a series of stages,
beginning with a market or sector that is successful, with strong
fundamentals. Credit expands, and money flows more easily. (Near the
peak of Japan's bubble in 1990, Japan's banks were lending money for
real estate purchases at more than the value of the property, expecting
the value to rise quickly.) As more money is available, prices rise.
More investors are drawn in, and expectations for quick profits rise.
The bubble expands, and then finally has to burst. In other words, fuelled by
initially well-founded economic fundamentals, investors develop a
self-fulfilling enthusiasm by an imitative process or crowd behavior
that leads to the building of castles in the air, to paraphrase Malkiel
(2012). Furthermore, the causes of the crashes on the US markets in
1929, 1987, 1998 and in 2000 belong to the same category, the
difference being mainly in which sector the bubble was created: in 1929,
it was utilities; in 1987, the bubble was supported by a general
deregulation of the market with many new private investors entering it
with very high expectations with respect to the profit they would
make; in 1998, it was an enormous expectation with respect to the
investment opportunities in Russia that collapsed; before 2000, it was
extremely high expectations with respect to the Internet,
telecommunications, and so on, that fuelled the bubble. In 1929, 1987 and
2000, the concept of a ``new economy'' was each time promoted as the
rational origin of the upsurge of the prices.

Several previous works in
economics have suggested that bubbles and crashes have endogenous origins, 
as we explain below. For instance, Irving Fisher (1933)
and Hyman Minsky (1992) both suggested that endogenous feedback effects lead to
financial instabilities, although their analysis did not include formal models.
Robert Shiller (2006) has been spearheading the notion that markets, at times, exhibit 
``irrational exuberance''. While the efficient market hypothesis provides a useful first-order 
representation of financial markets in normal times, one can observe regimes 
where the anchor of a fundamental price is shaky and large uncertainties characterize the future gains,
which provides a fertile environment for the occurrence of bubbles. When a number of additional elements are present, markets go through transient phases where they disconnect in specific dangerous ways from this 
fuzzy concept of fundamental value. These are regimes where investors are herding, following the flock and pushing the price up along an unsustainable growth trajectory.  
Many other mechanisms have been studied to explain the occurrence
of financial bubbles, such as constraints on short selling and lack of synchronisation
of arbitrageurs due to heterogeneous beliefs on the existence of a bubble,
see  Brunnermeier and Oehmke (2012) and Xiong (2013) for two excellent reviews.

\subsection{The critical point analogy}

Mathematically, we propose that large stock market crashes are the social analogues of
so-called critical points studied in the statistical physics community
in relation to magnetism, melting, and other phase transformation of
solids, liquids, gas and other phases of matter (Sornette, 2000). This
theory is based on the existence of a cooperative behavior of traders
imitating each other which leads to progressively increasing build-up of
market cooperativity, or effective interactions between investors, often
translated into accelerating ascent of the market price over months and
years before the crash. According to this theory, a crash occurs because
the market has entered an unstable phase and any small disturbance or
process may have triggered the instability. Think of a ruler held up
vertically on your finger: this very unstable position will lead
eventually to its collapse, as a result of a small (or absence of
adequate) motion of your hand or due to any tiny whiff of air. The
collapse is fundamentally due to the unstable position; the
instantaneous cause of the collapse is secondary. In the same vein, the
growth of the sensitivity and the growing instability of the market
close to such a critical point might explain why attempts to unravel the
local origin of the crash have been so diverse. Essentially, anything
would work once the system is ripe. In this view, a crash has
fundamentally an endogenous or internal origin and exogenous or external
shocks only serve as triggering factors.

As a consequence, the origin of crashes is much more subtle than often
thought, as it is constructed progressively by the market as a whole, as
a self-organizing process. In this sense, the true cause of a crash
could be termed a systemic instability. This leads to the possibility
that the market anticipates the crash in a subtle self-organized and
cooperative fashion, hence releasing precursory ``fingerprints''
observable in the stock market prices (Sornette and Johansen, 2001;
Sornette, 2003). These fingerprints have been modeled by
log-periodic power laws (LPPL) (Johansen et al., 1999; 2000), which are beautiful mathematical
patterns associated with the mathematical generalization of the notion
of fractals to complex imaginary dimensions (Sornette, 1998). In the framework of 
Johansen, Ledoit and Sornette (1999; 2000), 
an Ising-like stochastic dynamics is used to describe the time evolution of 
imitation between noise traders, which controls
the dynamics of the crash hazard rate (see Sornette et al. (2013) for a recent
update on the status of the model). 

Our theory of collective behaviour predicts robust signatures of
speculative phases of financial markets, both in accelerating bubbles
and decreasing prices (see below). These precursory patterns have been
documented for essentially all crashes on developed as well as emergent
stock markets. Accordingly, the crash of October 1987 is not unique but
a representative of an important class of market behaviour, underlying
also the crash of October 1929 (Galbraith, 1997) and many others
(Kindleberger, 2000; Sornette, 2003).

We refer to the book (Sornette, 2003) for a detailed description and the
review of many empirical tests and of several forward predictions. In
particular, we predicted in January 1999 that Japan's Nikkei index would
rise 50 percent by the end of that year, at a time when other economic
forecasters expected the Nikkei to continue to fall, and when Japan's
economic indicators were declining. The Nikkei rose more than 49 percent
during that time. We also successfully predicted several short-term
changes of trends in the US market and in the Nikkei and we have
diagnosed ex-ante several other major bubbles (see e.g. (Jiang et al., 2010)
and references therein).

\subsection{Tests with the financial crisis observatory}

In 2008, we created the Financial Crisis Observatory (FCO: http://www.er.ethz.ch/fco) as a scientific platform 
that aimed at testing and quantifying rigorously, in a systematic way and on 
a large scale the hypothesis that financial markets exhibit a degree 
of inefficiency and a potential for predictability, especially during regimes when bubbles develop.
Because back-testing is subjected to a host of possible biases, in November 2009, 
the Financial Bubble Experiment  (FBE) was launched within 
the FCO at ETH Zurich. Our motivation is to develop real-time advanced forecast
methodology that is constructed to be free, as much as possible, of all possible biases 
plaguing previous tests of bubbles.

In particular, active researchers are constantly tweaking their procedures, so that predicted `events' become moving targets. Only advanced forecasts can be free of data-snooping and other statistical biases of ex-post tests. The FBE aims at rigorously testing bubble predictability using methods developed in our group and by other scholars over the last decade. The main concepts and techniques used for the FBE have been documented in numerous papers (Jiang et al., 2009; Johansen et al., 1999; Johansen and Sornette, 2006; Sornette and Johansen, 2001; Sornette and Zhou, 2006b) and the book (Sornette, 2003). 
In the FBE, we developed a new method of delivering our forecasts where the results 
are revealed only after the predicted event has passed but where the original date when we produced these same 
results can be publicly, digitally authenticated (see
the reports and ex-post analysis of our diagnostics performed ex-ante at 
http://www.er.ethz.ch/fco and resources therein).

Stock market crashes are often unforeseen by most people, especially
economists. One reason why predicting complex systems is difficult is
that we have to look at the forest rather than the trees, and almost
nobody does that. Our approach tries to avoid this trap. From the tulip
mania, where tulips worth tens of thousands of dollars in present U.S.
dollars became worthless a few months later, to the U.S. bubble in 2000,
the same patterns occur over the centuries. Today we have electronic
commerce, but fear and greed remain the same. Humans remain endowed with
basically the same qualities (fear, greed, hope, lust)  today as they were in the 17th century.

\subsection{The social bubble hypothesis}

Bubbles and crashes are ubiquitous to human activity: we as
humans are rarely satisfied with the Status Quo; we tend to be
over-optimistic with respect to future prospects and, as social animals,
we herd to find comfort in being (right or wrong) with the crowd. This
leads to human activities being punctuated by bubbles and their
corrections. The bubbles may come as a result of expectations of the
future returns from new technology, such as in the exploration of the
solar system, of the human biology or new computer and information
technologies. I contend that this trait allows us as a species to take
risks to innovate with extraordinary successes that would not arise
otherwise. 

Bubbles defined as collective over-enthusiasm seem a necessary 
(and unavoidable) process to foster our collective attitude towards 
risk taking, breaking the stalemate of society resulting from 
its tendency towards strong risk avoidance (Sornette, 2008). An absence of bubble psychology 
would lead to stagnation and conservatism as no large risks are 
taken and, as a consequence, no large return can be accrued. We have coined 
the term `social bubble' in order to show how to take advantage of the bubble 
process to catalyze long-term investment (Sornette, 2008; Gisler and Sornette, 2009; 2010;
Gisler et al., 2011). A similar conclusion has been reached by
William Janeway (2012), an American venture capital investor for more than 40 years.
His book provides an accessible pathway to appreciate the dynamics of the innovation 
economy. In his understanding, informed by both practice and theory, the innovation economy begins with discovery and culminates in speculation, with continuous positive feedbacks loops between them. Over some 250 years, so his argument goes, economic growth has been driven by successive processes of trial and error: upstream explorations in research and inventions and downstream experiments in exploiting the new economic space opened by innovation. 

In a nutshell, the ``social bubble hypothesis'' claims that strong social interactions 
between enthusiastic supporters weave a network of reinforcing feedbacks 
that lead to widespread endorsement and extraordinary commitment by those involved, 
beyond what would be rationalized by a standard cost-benefit analysis. 
It does not cast any value system however, notwithstanding the use of the term ``bubble''.
Rather it identifies the types of dynamics that shape scientific or technological endeavors. 
In other words, we suggest that major projects often proceed via a social bubble mechanism
(Sornette, 2008; Gisler and Sornette, 2009; 2010; Gisler et al., 2011; 2013).

Thus, bubbles and crashes, the hallmark of humans, are
perhaps our most constructive collective process. But they may also
undermine our quest for stability. We thus have to be prepared and
adapted to the systemic instabilities that are part of us, part of our
collective organization, ... and which will no doubt recur again perhaps
with even more violent effects in the coming decade.

\section{Agent-based models (ABMs) in economics and finance}

Our review would be incomplete if not covering the very dynamical
field of agent based models (ABMs), also known as computational economic models.
They provide an alternative to the econometric and 
dynamical stochastic general equilibrium (DSGE) approaches
used by central banks for instance. They use computer simulated 
interactions between agents (decision-makers) (Farmer and Foley, 2009). The Ising-type models 
discussed in the preceding sections can be considered as special ABM implementations.

ABMs also illustrate vividly the special relations between economics and physics.
Consider Schelling (1971; 1978)'s work that demonstrated how slight 
differences of micro motives among heterogenous agents lead
to impressive macro behaviours. Schelling wanted to falsify the standard view
about segregations between black and white communities in the US, which 
assumed strong differences in preferences in order to explain the 
observed concentrations. Using manually implemented ABMs on 
a check board, he showed that tiny variations in tastes are sufficient
to lead to macroscopic segregation when allowing the system to evolve
over sufficiently long times. Small micro-effects lead to large macro-consequences.
This discovery was a breakthrough in the social sciences and changed
the perspective on community segregation.
To the physicist trained in the field
of phase transitions and statistical physics, this result is pretty obvious:
tiny differences in the interactions between pairs of molecules (oil-oil, water-water
and oil-water) are well-known to renormalise into macroscopic demixing. 
This is a beautiful example of the impact of repeated interactions 
leading to large-scale collective patterns. In the physicist language,
in addition to energy, entropy is an important and often leading contribution to large scale
pattern formation, and this understanding requires the typical statistical
physics training that economists and social scientists often lack.

\subsection{A taste of ABMs \label{abmliste}}

Agent-based models have the advantage of facilitating interdisciplinary collaboration and reveal unity across disciplines (Axelrod, 2005; Parisi et al., 2013). The possibilities of such models are a priori almost endless, only limited by the available computational power as well as the insights of the modeller. One can simulate very large number of different agents acting (up to tens of millions of agents as for instance in www.matsim.org, see (Meister et al., 2010) and other references at this url). Different decision making rules can be implemented, including utility maximization or behavioral decision making. For example, one can have different agents to model consumers, policy-makers, traders or institutions where each type follows possibly distinct agendas and obeys different decision making rules. Such a simulation is performed in discrete time steps where, at every time step, each actor has to take a decision (e.g. buying, selling or staying out of a stock on the financial market) based on her behavioral rules. Phenomena such as bubbles and subsequent crashes have been found to emerge rather naturally from such ABMs as a consequence of the existence of general interaction effects among heterogeneous agents. These interactions range from social herding, rational imitation to information cascades (Bikhchandani et al., 1992).

To study large-scale phenomena arising from micro-interactions, ABMs have already found numerous applications in the past (Bonabeau, 2002; MacKinzie, 2002). Early ABM developed for social science 
applications include F\"ollmer (1974)'s mathematical treatment of Ising economies with no stabilization for strong agent interactions, Schelling (1978)Õs segregation model, Weidlich (1991)'s synergetic approach, Kirman (1991, 1993)'s ant model of recruitment and so on. 

The Santa Fe Institute Artificial Stock Market is one of the pioneering ABMs, which was created by a group of economists and computer scientists at the Santa Fe Institute in New Mexico (Arthur et al., 1997; LeBaron et al, 1999; Palmer et al., 1994; 1999). The key motivation was to test whether artificially intelligent agents would converge to the homogeneous rational expectations equilibrium or not. To the surprise of the creators, the artificial stock markets failed to show convergence to the expected equilibrium, but rather underlined the importance of co-evolution of trading strategies adopted by the synthetic agents together with the aggregate market behavior. However, the Santa Fe Institute Artificial Stock Market has been shown to suffer from a number of defects, for instance the fact that the rate of appearance of new trading strategies is too fast to be realistic. Only recently was it also realized that previous interpretations neglecting the emergence of technical trading rules should be corrected (Ehrentreich, 2008). 

Inspired by the El-Farol Bar problem (Arthur, 1994b) meant to emphasize how inductive reasoning together with a ÒminorityÓ payoff prevents agents to converge to an equilibrium and force them to continuously readjust their expectation, the Minority Game was introduced by Challet and Zhang (1997; 1998) to model prices in markets as reflecting competition among a finite number of agents for a scarce resource (Marsili et al., 2000).  Extensions include the Majority Game and the Dollar Game (a time delayed version of the majority game) and delayed version of the minority games. In minority games, which are part of first-entry games, no strategy can remain persistently a winner; otherwise it will be progressively adopted by a growing number of agents, bringing its demise by construction of the minority payoff. This leads to the phenomenon of ÒfrustrationÓ and anti-persistence. Satinover and Sornette (2007a,b; 2009) have shown that optimizing agents are actually underperforming random agents, thus embodying the general notion of the Òillusion of controlÓ. It can be shown more generally that learning and adaptive agents will converge to the best dominating strategy, which turns out to be the random choice strategy for minority or first-entry payoffs. 

Evstigneev et al. (2009) review results obtained on evolutionary finance, namely the field studying the dynamic interaction of investment strategies in financial markets through ABM implementing Darwinian ideas and random dynamical system theory. By studying the wealth distribution among agents over the long-term, Evstigneev et al. are able to determine the type of strategies that over-perform in the long term. They find that such strategies are essentially derived from Kelly (1956)'s criterion of optimizing the expected log-return. They also pave the road for the development of a generalization of continuous-time finance with evolutionary and game theoretical components. 

Darley and Outkin (2007) describe the development of a Nasdaq ABM market simulation, developed during the collaboration between the Bios Group (a spin-off of the Santa Fe Institute) and Nasdaq Company to explore new ways to better understand Nasdaq's operating world. The artificial market has opened the possibility to explore the impact of market microstructure and market rules on the behavior of market makers and traders. One obtained insight is that decreasing the tick size to very small values may hinder the market's ability to perform its price discovery process, while at the same time the total volume traded can greatly increase with no apparent benefits (and perhaps direct harm) to the investors' average wealth. 

In a similar spirit of using ABM for an understanding of real-life economic developments, Geanakoplos et al. (2012) have developed an agent based model to describe the dynamics that led to the housing bubble in the US that peaked in 2006 (Zhou and Sornette, 2006). At every time step, the agents have the choice to pay a monthly coupon or to pay off the remaining balance (prepay). The conventional method makes a guess for the functional form of the prepayments over time, which basically boils down to extrapolate into the future past patterns in the data. In contrast, the ABM takes into account the heterogeneity of the agents through a parameterization with two variables that are specific to each agent: the cost of prepaying the mortgage and the alertness to his financial situation. A simulation of such agents acting in the housing market is able to capture the run up in housing price and the subsequent crash. The dominating factor driving this dynamic could be identified as the leverage the agents get from easily accessible mortgages. The conventional model entirely missed this dynamic and was therefore unable to forecast the bust. Of course, this does not mean that non-ABM models have not been able or would not be able to develop the insight about the important role of the procyclicality of the leverage on real-estate prices and vice-versa, a mechanism that has been well and repeatedly described in the literature after the crisis in 2007-2008 erupted.  

Hommes (2006) provides an early survey on dynamic behavioral financial and economic models with rational agents with bounded rationality using different heuristics. He emphases the class of relatively simple models for which some tractability is obtained by using analytic methods in combination with computational tools. Nonlinear structures often lead to chaotic dynamics, far from an equilibrium point, in which regime switching is the natural occurrence associated with coexisting attractors in the presence of stochasticity (Yukalov et al., 2009). By the aggregation of relatively simple interactions occurring at the micro level, quite sophisticated structure at the macro level may emerge, providing explanations for observed stylized facts in financial time series, such as excess volatility, high trading volume, temporary bubbles and trend following, sudden crashes and mean reversion, clustered volatility and fat tails in the returns distribution.

Chiarella et al. (2009) review another branch of investigation of boundedly rational heterogeneous agent models of financial markets, with particular emphasis to the role of the market clearing mechanism, the utility function of the investors, the interaction of price and wealth dynamics, portfolio implications, and the impact of stochastic elements on market dynamics. Chiarella et al. find regimes with market instabilities and stochastic bifurcations, leading to fat tails, volatility clustering, large excursions from the fundamental, and bubbles, which are features of real markets that are not easily reconcilable for the standard financial market paradigm.

Shiozawa et al., 2008 summarize the main properties and finding resulting from the U-Mart project, which creates a virtual futures market on a stock index using a computer or network in order to promote on-site training, education, and economics research. In the U-Mart platform, human players can interact with algorithms, providing a rich experimental platform. 

Building on the insight that when past information is limited to a rolling window of prior states of fixed length, the minority, majority and dollar games may all be expressed in Markov-chain formulation (Marsili et al. 2000, Hart et al. 2002, Satinover and Sornette 2007a,b), Satinover and Sornette (2012a,b) have further shown how, for minority and majority games, a cycle decomposition method allows to quantify the inherently probabilistic nature of a Markov chain underlying the dynamics of the models as an exact superposition of deterministic cyclic sequences (Hamiltonian cycles on binary graphs), extending ideas discussed by Jefferies et al. (2002). This provides a novel technique to decompose the sign of the time-series they generate (analogous to a market price time-series) into a superposition of weighted Hamiltonian cycles on graphs. The cycle decomposition also provides a ÔdissectionÕ of the internal dynamics of the games and a quantitative measure of the degree of determinism. The performance of different classes of strategies may be understood on a cycle-by-cycle basis. The decomposition offers a new metric for comparing different game dynamics with real-world financial time-series and a method for generating predictors. A cycle predictor applied to a real-world market can generate significantly positive returns.

Feng et al. (2012) use an agent-based model that suggests a dominant role for the investors using technical strategies over those with fundamental investment styles, showing that herding emerges via the mechanism of converging on similar technical rules and trading styles in creating the well-known excess volatility phenomenon (Shiller, 1981; LeRoy and Porter, 1981; LeRoy, 2008). This suggests that there is more to price dynamics than just exogeneity (e.g. the dynamics of dividends). Samanidou et al. (2007) review several agent-based models of financial markets, which have been studied by economists and physicists over the last decade: Kim-Markowitz, Levy-Levy-Solomon (1994), Cont-Bouchaud, Solomon-Weisbuch, Lux-Marchesi (1999; 2000), Donangelo-Sneppen and Solomon-Levy-Huang. These ABM emphasize the importance of heterogeneity, of Ònoise tradersÓ (Black, 1986) or technical analysis based investment styles, and of herding. Lux (2009a) reviews simple stochastic models of interacting traders, whose design is closer in spirit to models of multiparticle interaction in physics than to traditional asset-pricing models, reflecting the insight emergent properties at the macroscopic level are often independent of the microscopic details of the system. Hasanhodzic et al. (2011) provides a computational view of market efficiency, by implementing agent-based models in which agents with different resources (e.g, memories) perform differently. This approach is very promising to understand the relative nature of market efficiency (relative with respect to resources such as super-computer power and intellectual capital) and provides a rationalization of the technological arm race of quantitative trading firms.

\subsection{Outstanding open problems: robustness and calibration/validation of agent-based models}

The above short review gives a positive impression on the potential of ABMs. In fact, orthodox (neoclassical) economists have in a sense taken stock of the advances provided by ABMs by extending their models to include ingredients of heterogeneity, bounded rationality, learning, increasing returns, and technological change. Why then are not ABMs more pervasive in the work of economists and in the process of decision making in central banks and regulators? We think that there are two dimensions to this question, which are interconnected (see also Windrum et al, 2007).  

First, ABMs have the disadvantage of being complicated with strong nonlinearities and stochasticity in the individual behaviors, made of multiple components connected through complex interactive networks, and it is often difficult to relate the resulting outcomes from the constituting ingredients. In addition, the feedbacks between the micro and macro levels lead to complex behavior that cannot be analyzed analytically, for instance by the powerful tool of the renormalization group theory (Wilson, 1979; Goldenfeld, 1993; Cardy, 1996), which has been so successful in statistical physics in solving the micro-macro problem (Anderson, 1972; Sornette, 2004) by the flow of the change
of the descriptive equations of a complex system when analyzed at different resolution scales. The different types of agents and their associated decision making rules can be chosen without much restrictions to encompass the available knowledge in decision theory and behavioral economics. However, the choices made to build a given agent-based model may represent the personal preferences or biases of the modeler, which would not be agreeable to another modeler. ABMs are often constructed with the goal of illustrating a given behavior, which is actually already encoded more or less explicitly in the chosen rules (De Grauwe, 2010; Galla and Farmer, 2013). Therefore, the correctness of the model relies mostly on the relevance of the used rules, and the predictive power is often constrained to a particular domain so that generalization is not obvious. This makes it difficult to compare the different ABMs found in the literature and gives an impression of lack of robustness in the results that are often sensitive to details of the modelerÕs choices. The situation is somewhat similar to that found with artificial neural network, the computational models inspired by animals' central nervous systems that are capable of machine learning and pattern recognition. While providing interesting performance, artificial neural networks are Òblack boxesÓ: it is generally very difficult if not impossible to extract a qualitative understanding of the reasons for their performance and ability to solve a specific task. We can summarize this first difficulty as the Òmicro-macroÓ problem, namely understanding how micro-ingredients and rules transform into macro-behaviors at the collective level when aggregated over many agents.

The second related problem is that of calibration and validation (Sornette et al., 2007). Standard DSGE models of an economy, for instance, provide specific regression relations that are relatively easy to calibrate to a cross-sectional set of data. In contrast, the general problem of calibrating ABMs is unsolved. By calibrating, we refer to the problem of determining the values of the parameters (and their uncertainty intervals) that enter in the definition of the ABM, which best corresponds to a given set of empirical data. Due to the existence of nonlinear chaotic or more complex dynamical behaviors, the likelihood function is in general very difficult if not impossible to determine and standard statistical methods (maximum likelihood estimation (MLE)) cannot apply. Moreover, due to the large number of parameters present in large scale ABMs, calibration suffers from the curse of dimensionality and of ill-conditioning: small errors in the empirical data can be amplified into large errors in the calibrated parameters. We think that it is not exaggerated to state that the major obstacle for the general adoption of ABMs by economists and policy makers is the absence of a solid theoretical foundation for and efficient reliable operational calibration methods.  

This diagnostic does not mean that there have not been attempts, sometime quite successful, in calibrating ABMs. Windrum et al. (2007) review the advances and discuss the methodological problems arising in the empirical calibration and validation of agent-based models in economics. They classify the calibration methods into three broad classes: (i) the indirect calibration approach, (ii) the Werker-Brenner approach, and (iii) the history-friendly. They have also identified six main methodological and operational issues with ABM calibration: (1) fitness does not imply necessarily that the true generating process has been correctly identified; (2) the quest for feasible calibration influences the type of ABMs that are constructed; (3) the quality of the available empirical data; (4) the possible non-ergodicity of the real world generating process and the issue of representativeness of short historical time series; (5) possible time-dependence of the micro and macro parameters.

Restricting our attention to financial markets, an early effort of ABM calibration is that of Poggio et al. (2001), who constructed a computer simulation of a repeated double-auction market. Using six different experimental designs, the calibration was of the ÒindirectÓ type, with attempt to match the price efficiency of the market, the speed at which prices converge to the rational expectations equilibrium price, the dynamics of the distribution of wealth among the different types of artificial Intelligent agents, trading volume, bid/ask spreads, and other aspects of market dynamics. Among the ABM studies touched upon above, that of Chiarella et al. (2009) include an implementation of the indirect calibration approach. Similarly, Bianchi et al. (2007) develop a methodology to calibrate the Complex Adaptive Trivial System (CATS) model proposed by Gallegati et al. (2005), again matching several statistical outputs associated with different stylized facts of the ABM to the empirical data. Fabretti (2013) uses a combination of mean and standard deviation, kurtosis, Kolmogorov-Smirnov statistics and Hurst exponent for the statistical objects determined from the ABM developed by Farmer and Joshi (2002) whose distance to the real statistics should be minimized. 

Alfarano et al. (2005) studied a very simple ABM that reproduces the most studied stylized facts (fat tails, volatility clustering). The simplicity of the model allows the authors to derive a closed form solution for the distribution of returns and hence to develop a rigorous maximum likelihood estimation (MLE) approach to the calibration of the ABM. The analytical analysis provides an explicit link between the exponent of the unconditional power law distribution of returns and some structural parameters, such as the herding propensity and the autonomous switching tendency. This is a rare example for which the calibration of the ABM is similar to more standard problems of calibration in econometrics.

Andersen and Sornette (2005) introduced a direct Òhistory-friendlyÓ calibration method of the Minority game on time series of financial returns, which statistically significant abnormal performance to detect special Òpockets of predictabilityÓ associated with turning points. Roughly speaking, this is done by calibrating many times the ABM to the data and by performing meta-searches in the set of parameters and strategies, while imposing robustness constraints to address the intrinsic ill-conditional nature of the problem. One of the advantages is to remove possible biases of the modeler (except for the fact that the structure of the model reflects itself a view of what should be the structure of the market). This work by Andersen and Sornette (2005) was one of the first to establish the existence of pockets of predictability in stock markets. A theoretical analysis showed that when a majority of agents follows a decoupled strategy, namely the immediate future has no impact on the longer-term choice of the agents, a transient predictable aggregate move of the market occurs. It has been possible to estimate the frequency of such prediction days if the strategies and histories were randomly chosen. A statistical test, using the Nasdaq Composite Index as a proxy for the price history, confirms that it is possible to find prediction days with a probability extremely higher then chance.

Another interesting application is to use the ABM to issue forecasts that are used to further refine the calibration as well as test the predictive power of the model. To achieve this, the strategies of the agents become in a certain sense a variable, which is optimized to obtain the best possible calibration of the in-sample data. Once the optimal strategies are identified, the predictive power of the simulation can be tested on the out-of-sample data. Statistical tests have shown that the model performs significantly better than a set of random strategies used as comparison (Andersen and Sornette; 2005; Wiesinger et al., 2012). These results are highly relevant, because they show that it seems possible to extract from times series information about the future development of the series using the highly nonlinear structure of ABMs. Applied to financial return time series, the calibration and subsequent forecast show that the highly liquid financial markets (e.g. S\&P500 index) have progressively evolved towards better efficiency from the 1970s to present (Wiesenger et al., 2012). Nevertheless, there seems to remain statistically significant arbitrage opportunities (Zhang, 2013), which seems inconsistent with the weak form of the efficient market hypothesis (EMH). This method lays down the path to a novel class of statistical falsification of the EMH. As the method is quite generic, it can virtually be applied on any time series to check how well the EMH holds from the viewpoint offered by the ABM. Further on, this approach has wide potential to reverse engineer many more stylized facts observed in financial markets.

Lillo et al. (2008) present results obtained in the rare favorable situation in which the empirical data is plentiful, with access to a comprehensive description of the strategies followed by the firms that are members of the Spanish Stock Exchange. This provides a rather unique opportunity for validating the assumptions about agentÕs preferred stylized strategies in ABMs. The analysis indicates that three well-defined groups of agents (firms) characterize the stock exchange.

Saskia Ter and Zwinkels (2010) have modeled the oil price dynamics with a heterogeneous agent model that, as in many other ABMs, incorporates two types of investors, the fundamentalists and the chartists and their relation to the fundamental supply and demand. The fundamentalists, who expect the oil price to move towards the fundamental price, have a stabilizing effect, while the chartists have a destabilizing effect driving the oil price away from its fundamental value. The ABM has been able to outperform in an out-of-sample test both the random walk model and VAR models for the Brent and WTI market, providing a kind of partial Òhistory-friendlyÓ calibration approach.

\subsection{The ``Emerging Market Intelligence Hypothesis'' \label{gsnewh}}

Financial markets can be considered as the engines that transform information into price.
The efficient market hypothesis (EMH) states that the continuous efforts
of diligent investors aggregate into a price dynamics that does not contain any arbitrage 
opportunities (Samuelson, 1965; 1973; Fama, 1970; 1991). In other words, the very process of 
using better information or new technology to 
invest with the goal of generating profits in excess to the long-term historical market growth rate 
makes the prices unfathomable and destroys the very goals of the investors. 

Farmer (2002) constructed simple set-ups in which the mean-reversion nature
of investors' strategies stabilise prices and tends to remove arbitrage opportunities.
Satinover and Sornette (2007a;b; 2009) showed how the ``whitening''
(i.e. destruction of any predictive patterns) of the prices precisely occur
in minority games (Challet and Zhang, 1998; 1999; Challet et al., 2005).
Specifically,  agents who optimize their strategy based on available information
actually perform worse than non-optimizing agents. In other words, low-entropy (more informative)
strategies under-perform high-entropy (or random) strategies. This results from 
an emergent property of the whole game that no non-random strategy can
outwit. Minority games can be considered as subset of first-entry games, for which the same
phenomenon holds (Duffy and Hopkins, 2005). 
In first-entry games, this means that agents who learn on stochastic fictitious plays
will adapt and modify their strategies to finally converge to the 
best strategies, which randomize over the entry decisions. 
Thus, in minority and first-entry games, when players think that they can put some sense
to the patterns created by the games,
that they have found a winning strategy and they have an advantage, they are
delusional since the true winning strategies are random. 

In reality, efficient markets do not exist. 
Grossman and Stiglitz (1980) articulated in a simplified model the essence 
of a quite intuitive mechanism: 
because gathering information is costly, prices cannot perfectly reflect all the information that is available since this would confer no competitive advantage to those who spent resources to obtain it and trade on it, therefore destroying the very mechanism by which information is incorporated into prices. As a consequence, an informationally efficient market is impossible and the Efficient Market Hypothesis (EMH) can only be a first-order approximation, an asymptotic ideal construct that is never reached in practice.
It can be approached but a convergence towards it unleash effective repelling forces due to dwindling incentives. ``The abnormal returns always exist to compensate for the costs of gathering and processing information. These returns are necessary to compensate investors for their information-gathering and information-processing expenses, and are no longer abnormal when these expenses are properly accounted for. The profits earned by the industrious investors gathering information may be viewed as economic rents that accrue to those willing to engage in such activities'' (cited from Campbell et al., 1997).

Let us push this reasoning in order to illuminate further the nature and limits of the EMH and as a bonus clarify the nature and origin of ``noise traders'' (Black, 1986). As illustrated by the short review of
section \ref{abmliste}, the concept of ``noise trader'' is 
an essential constituent of most ABMs that aim at explaining the excess volatility, fat-tailed distributions
of asset returns, as well as the astonishing occurrence of bubbles and crashes. It also solves
the problem of the no-trade theorem (Milgrom and Stokey, 1982), which in essence
shows that no investor will be willing to trade if the market is 
in a state of efficient equilibrium and there are no 
noise traders or other non-rational interferences with prices. Intuitively,
if there is a well-defined fundamental value, all well-informed rational traders
agree on it, the market prices is the fundamental value and 
everybody holds the stock according to their portfolio allocation strategy
reflecting their risk profiles. No trade is possible without the existence
of exogenous shocks, changes of fundamental values or taste alterations.

In reality, real financial markets
are heavily traded, with at each tick an exact balance between the total
volume of buyers and of sellers (by definition of each realised trade), reflecting a generalised
disagreement on the opportunity to hold the corresponding stocks. These
many investors who agree to trade and who trade much more than would 
be warranted on the basis of fundamental information are called noise traders.
Noise traders are loosely defined as the investors 
who makes decisions regarding buy and sell trades without much use of fundamental data, 
but rather on the basis of price patterns, trends, and who react incorrectly
to good and bad news. On one side, traders exhibit over-reaction, which refers to the 
phenomenon that price responses to news events are exaggerated.
A proposed explanation is that excess price pressure is applied by 
overconfident investors (Bondt and Thaler, 1985; Daniel et al., 1998) 
or momentum traders (Hong and Stein, 1999),
resulting in an over- or under-valued asset, which then
increases the likelihood of a rebound and thus creates a negative autocorrelation in returns. 
On the other side, investors may under-react, resulting in 
a slow internalisation of news into price. Due to such temporally spread-out impact of the news, 
price dynamics exhibit momentum, i.e. positive return autocorrelation
(Lo and MacKinlay, 1988; Jegadeesh, 1990; Cutler, 1990; Jegadeesh and Titman, 1993).

In fact, most investors and portfolio managers are considered noise traders (Malkiel, 2012)!
In other words, after controlling for luck, there is a general consensus in the
financial academic literature that most fund managers do not provide
statistically significant positive returns above the market return that would be obtained
by just buying and holding for the long term (Barras et al., 2010; Fama and French, 2010).
This prevalence of noise traders is in accord with the EMH. But are these investors
really irrational and mindless? This seems difficult to reconcile with the evidence
that the banking and investment industry has been able in the last decades
to attract a significant fraction of the best minds and most motivated persons on Earth.
Many have noticed and even complained that, in the years before the financial crisis of 2008, the best
and brightest college grads were heading for Wall Street.
At ETH Zurich where I teach financial market risks and tutor master theses, I have observed
even after the financial crisis a growing flood of civil, mechanical,  electrical and other engineers 
choosing to defect from their field and work in finance and banking.  

Consequently, we propose that noise traders are actually highly intelligent motivated and capable 
investors. They are like noise traders as a result of the aggregation of the collective intelligence 
of all trading strategies that structure the price dynamics and makes each of the individual strategy look
``stupid'', like noise trading. The whole is more than the sum of the part. 
 In other words, a universe of rational optimizing traders create endogenously 
 a large fraction of rational traders who are effectively noise, because their strategies are like noise, given the complexity or structure of financial and economic markets that they collectively create. 
 The continuous actions of investors, which are aggregated in the prices, produce a ``market intelligence'' more powerful than that of most of them. The ``collective intelligence'' of the market transforms most (but not all) strategies into 
losing strategies, just providing liquidity and transaction volume. 
We call this the  ``Emerging Market Intelligence hypothesis'' (EIMH). This phrasing stresses the collective intelligence that dwarfs the individual ones, making them look like noise when applied 
 to the price structures resulting from the price formation process.
 
But for this EIMH to hold, the ``noise traders'' need a motivation to continue trading, 
in the face of their collective dismal performances. In addition to 
the role of monetary incentives for rent-seeking that permeates the banking industry (Freeman, 2010)
and makes working in finance very attractive notwithstanding the absence of 
genuine performance, there is a well-documented fact in the field of psychology
that human beings in general and investors in particular (and especially traders who are (self-)selected for their distinct abilities and psychological traits) tend to rate their skills over-optimistically (Kruger and Dunning, 1999).
And when by chance, some performance emerges, we tend to attribute the positive outcome
to our skills. When a negative outcome occurs, this is bad luck. This is referred to in the
psychological literature as ``illusion of control'' (Langer 1975). 
In addition, human beings  have evolved the ability to attribute
meaning and regularity when there is none. In the psychological
literature, this is related to the fallacy of ``hasty generalisation''
(``law of small numbers'') and to ``retrospective determinism'', which makes
us look at historical events as part of an unavoidable meaningful laminar flow.
All these elements combine to generate a favourable environment to catalyse
trading, by luring especially young bright graduate students to finance in the belief
that their intelligence and technical skills will allow them to ``beat the market''. 
Thus, building on our cognitive biases and in particular on over-optimism, 
one could say that the incentive structures of the financial industry provides the
remunerations for the individuals who commit themselves to arbitrage the financial 
markets, thereby providing an almost efficient functioning machine.
The noise traders naturally emerge as a result of the emergent collective intelligence.
This concept is analogous to the sandpile model of self-organised criticality (Bak, 1996),
which consistency functions at the edge of chaos, driven to its instability but 
never completely reaching it by the triggering of avalanches (Scheinkman and Woodford, 1994). 
Similarly, the incentives of the financial system creates an army of highly
motivated and skilled traders who push the market towards efficiency but never
succeeding to allow some of them to win, while 
make most of them look like noise.

 \section{Concluding remarks}

While it is difficult to argue for a physics-based foundation
 of economics and finance, physics has still a role to play as a
 unifying framework full of concepts and tools to deal with complex
 dynamical out-of-equilibrium systems.
 Moreover, the specific training of physicists explains the impressive number of
 recruitments in investment and financial institutions, where their
 data-driven approach coupled with a pragmatic sense of theorizing has
 made physicists a most valuable commodity on Wall Street.
 
  At present however, the most exciting progress seems to be unraveling at the
 boundary between economics and the biological, cognitive and behavioral
 sciences (Camerer et al., 2003; Shiller, 2003; Thaler 2005).  
 A promising recent trend is the enrichment of financial economics
 by concepts developed in evolutionary biology. Several notable groups
 with very different backgrounds have touched upon the concept that 
 financial markets may be similar to ecologies filled by species
 that adapt and mutate. For instance,
 we mentioned earlier that Potters et al. (1998)
showed that the market has empirically corrected and adapted to the simple, but inadequate Black-Scholes formula  to account for the fat tails and the correlations in the scale of fluctuations. 
Doyne Farmer (2002) proposed a theory based on the interrelationships of
strategies, which views a market as a financial ecology. In this ecology, new
better adapted strategies exploit the inefficiencies of old strategies and 
the evolution of the capital of a strategy is analogous to the evolution 
of the population of a biological species.  
Cars Hommes (2001) also reviewed works modeling financial markets 
as evolutionary systems constituted of
different, competing trading strategies. Strategies are again taken
as the analog of species. It is found that simple technical
trading rules may survive evolutionary competition in a heterogeneous world
where prices and beliefs coevolve over time. Such evolutionary models
can explain most of the stylized facts of financial markets (Chakraborti et al., 2011).

Andrew Lo (2004; 2005; 2011) coined the term ``adaptive market hypothesis''
in reaction to the ``efficient market hypothesis'' (Fama, 1970; 1991), to propose an
evolutionary perspective on market dynamics in which intelligent but fallible 
investors learn from and adapt to changing environments, leading to 
a relationship between risk and expected return that is not constant in time.
In this view, markets are not always efficient but they are highly competitive and adaptive, 
and can vary in their degree of efficiency as the economic environment and 
investor population change over time.  Lo emphasizes that adaptation
in investment strategies (Neelya et al. 2009) are driven by the ``push for survival''. 
This is perhaps a correct assessment of Warren Buffet's own stated
strategy: ``We do  not wish it only to be likely
that we can meet our obligations; we wish that to be certain. Thus we adhere
to policies -- both in regard to debt and all other matters -- that will allow us
to achieve acceptable long-term results under extraordinary adverse conditions, 
rather than optimal results under a normal range of conditions'' 
(Berkshire Hathaway Annuel Report 1987: \url{http://www.berkshirehathaway.com/letters/1987.html}).
But the analogy with evolutionary biology as well as many studies of
the behavior of bankers and traders (e.g. Coates, 2012) suggest that most 
market investors care for much more than just survival. They strive to 
maximize their investment success measured as bonus and wealth, which
can accrue with luck on time scales of years. This is akin
to maximizing transmission of ``genes'' in a biological context (Dawkins, 1976).
The focus on survival within an evolutionary analogy is clearly insufficient
to account for the extraordinary large death rate of business companies and in particular
of financial firms such as hedge-funds (Saichev et al., 2010; Malevergne et al., 2013
and references therein). 

But evolutionary biology itself is witnessing a revolution with genomics, benefitting
from computerized automation and artificial intelligence classification
(ENCODE Project Consortium, 2012). And (bio-)physics
is bound to continue playing a growing role to organize the wealth of data
in models that can be handled, playing on the triplet of experimental, computational
and theoretical research. On the question of what could be useful tools
to help understand, use, diagnose, predict and control  financial markets 
(Cincotti et al., 20120; de S. Cavalcante et al. 2013), 
we envision that both physics and biology are going to play a growing role
to inspire models of financial markets, 
and the next significant advance will be obtained by marrying the three fields.

\end{document}